\documentclass[9.5pt,twocolumn]{article}

\usepackage[a4paper,margin=0.7in]{geometry}

\usepackage{amsmath,amssymb,amsthm}
\usepackage{graphicx}
\usepackage{booktabs}
\usepackage{multirow}
\usepackage{authblk}
\usepackage{natbib}

\title{Improved prediction of extreme random effects in joint models: WRaPs}

\author[1]{Eline Vanderpijpen}
\author[2]{Els Goetghebeur}

\affil[1,2]{Department of Mathematics, Computer Science and Statistics, Ghent University,
Ghent, Belgium.}

\date{June 17, 2026}

\def\bSig\mathbf{\Sigma}

\newcommand{\keywords}[1]{%
  \vspace{0.5em}
  \noindent\textbf{Keywords: }#1
}
\usepackage{fancyhdr}

\begin{document}
\twocolumn[
\maketitle
\begin{abstract}
Mixed models are popular for the prediction of subject-specific repeated outcomes or center performance among many centers.
When the goal is to identify extreme or poor outcomes, standard random effects predictions may, however, suffer from regression to the mean and underestimate values in the tail of their distribution. 
Optimally weighted random effect estimators have recently been proposed to mitigate this. 
Motivated by clinical settings where repeated outcomes may end in death, we extend that method to predict poor outcome defined as 'death or substandard repeated measures'. We start from joint models with shared random effects for the longitudinal and survival outcome and estimate their random effects by minimizing squared weighted prediction errors given available data on survival and repeated measures. As for mixed models, weights are chosen to more heavily penalize errors in the tails. We call the results WRaPs: Weighted Random effect Predictors. For basic models and a select set of weights analytical closed form solutions are derived from the usual joint model parameters. For the more complex setting, computational solutions are developed in rjags using MCMC methods within the Bayesian paradigm. We illustrate finite sample properties of the proposed method in Type I simulations with random intercept and slope; and apply the new approach to predict individual future outcomes and survival in a randomized study with glioblastoma patients.
\end{abstract}

\vspace{2em}
]
\section{Introduction}
\label{s:intro}

Personalized medicine and the identification of high-risk patients are becoming increasingly important in modern healthcare. 
Individual risk predictions are informing treatment  decisions in prostate cancer \citep {Taylor2005IndividualizedPredictions},
 can help improve long-term psychological and functional outcomes in oncology \citep{brown2023fear}, and support discussions 
in ICU to help  reduce disproportionate care arising from poor ethical decision-making \citep{porter2025discussing}.

Existing prediction tools in cancer trials primarily focus on event-based outcomes such as disease progression or survival. In an ICU context, with older patients or for aggressive cancers, however, limited survival benefits may not outweigh treatment-related side-effects. When a substantial survival increase may come with substantial loss of quality of life, one should consider patient preferences \citep{basch2015patient,beil2025understanding}.   A large consortium of diverse stakeholders recently emphasized the importance of  Patient Reported Outcomes and Quality of Life (QoL)  besides survival when evaluating treatments in oncology  \citep{Amdal2025}. Incorporating QoL allows to complement the patient-specific question phrased as “What are my chances of surviving?” with “If I survive, what quality of life  can I expect?”, shifting toward a two-dimensional outcome capturing both survival and QoL \citep{Reynders2025TwoDimensionalEstimand}.

These outcomes are typically correlated as poor health may come with poor Quality of Life and QoL tends to decline near death \citep{AhlnerElmqvist2009HRQOL}. 
This dependence can generate biased parameter estimates if conventional methods, such as linear mixed models, are applied to QoL data without accounting for death. Joint models solve this issue by simultaneously modeling longitudinal QoL and survival through shared random effects \citep{WulfsohnTsiatis1997JointModel}. They have been applied for individualized prediction, for instance in personalized biopsy scheduling \citep{tomer2019}, and are implemented within the \texttt{JM} package in \texttt{R} \citep{rizopoulos2010, rizopoulos2011}.

For individualized prediction in joint models with normal random effects, predicted random effects are often obtained as empirical Bayes estimates, which are analogous to Best Linear Unbiased Predictors (BLUPs) in linear mixed models \citep{henderson1975}. Since BLUPs tend to shrink extreme values toward the population mean \citep{robinson1991}, they may yield overly optimistic predictions for patients with very poor health or QoL. To handle this, \citet{mcculloch2021} minimized a weighted mean squared error of random effects in a linear mixed model, emphasizing the tails of the random effects distribution, to improve identification of extreme clusters \citep{mcculloch2024}. We extend this to the joint modeling context and derive Weighted Random effect Predictors (WRaPs), enhancing the flagging of individuals who are expected to die or survive with unacceptably poor quality of life over a predefined future period. Tuning the weight functions can balance true and false positive rates for more flexible control over prediction performance.

We apply our method analysing data from Glioblastoma patients in an EORTC trial, published before by \cite{Taphoorn2005HRQoL} and \cite{Stupp2005Radiotherapy}.

\section{Joint model framework}
Outcomes for person \(i\) \((i = 1, \dots,N)\) at time points \(t_{ij}\) \((j = 1, \dots, N_i)\) are denoted \(Y_{ij} = Y_i(t_{ij})\) and modeled by a linear mixed model:
\begin{eqnarray*}
    Y_{ij} = Y_i(t_{ij}) = X_{ij}'\beta + Z_{ij}'U_i + \epsilon_{ij}
\end{eqnarray*}
with $p$-dimensional vector of fixed effects \(\beta\), $q$-dimensional vector of subject-specific random effects (REs) \(U_i\), and residual error \(\epsilon_{ij}\). \(Z_{ij} \subset X_{ij}\) are vectors of possibly time-dependent covariates, fully determined by baseline information, the time variable \(t_{ij}\) and possible interactions with \(t_{ij}\). Without loss of generality $U_i \sim N(0, G)$, and $\epsilon_{ij} \sim N(0, \sigma_\epsilon^2)$, with the REs and residuals all jointly independent.

To account for dependence between the longitudinal and survival processes, our joint model links both processes through shared REs \citep{guo2004separate}:
\begin{eqnarray}\label{joint_model_eq}
    Y_{ij} &=&X_{ij}'\beta + Z_{ij}'U_i + \epsilon_{ij} \nonumber \\
    h_i(t|X_i(t),B_i(t),U_i) &=& h_i(t|B_i(t),U_i) \\&=& h_0(t) \exp(B'_i(t)\gamma_B + U'_i \gamma_u) \nonumber
\end{eqnarray}
involving baseline hazard $h_0(t)$ and a vector of possibly time-dependent covariates fully determined by baseline information $B_i$ with regression coefficient $\gamma_B$. Equation (\ref{joint_model_eq}) implies that \(B_i(t)\) covers parts of \(X_i(t)\) that remain relevant. Survival data enter as $D_i$, the observation time (minimum of the event time $T_i$ and censoring time $C_i$), and $\delta_i$, the event indicator \(I(T_i \leq C_i)\). We thus adopt the Full Conditional Independence Assumption, where REs $U_i$ account for the association between the longitudinal measurements and the event times, as well as repeated measurements correlation within subjects \citep{rizopoulos2010}:
\begin{eqnarray}\label{joint assumptions}
\left\{
\begin{array}{l}
    p(Y_i, D_i, \delta_i \mid U_i) = p(Y_i \mid U_i) p(D_i, \delta_i \mid U_i)\\
    p(Y_i \mid U_i) = \prod_j p(Y_{ij} \mid U_i)
    \end{array}
    \right.
\end{eqnarray}
For both processes, censoring is assumed to be non-informative conditional on the REs and included covariates. 
Estimated model parameters \((\beta,\gamma,G,\sigma_{\epsilon})\) can be obtained from maximum likelihood estimators (MLE) using the EM algorithm \citep{WulfsohnTsiatis1997JointModel,rizopoulos2009} or as the mean of the Bayesian posterior distribution \citep{guo2004separate,Rizopoulos2016JMbayes}.

\section{Prediction with joint model}
Within the 2D outcome framework, we focus on the subject specific survival function \( S(t|B_i(t), U_i) \) and the expected longitudinal outcome given survival at time \( t \), \( E\!\left(Y_i(t) \mid X_i(t), U_i, T_i > t\right) \). Under conditional independence of \(Y_i(t)\) and \(I(T_i>t)\) given \(X_i(t)\) and \(U_i\), the latter simplifies to \( E\!\left(Y_i(t) \mid X_i(t), U_i\right) \) \citep{rouanet2019}. Both quantities are predicted using the estimated joint model parameters and derived subject-specific predicted REs:
\begin{eqnarray}\label{JM_definition}
    \hat{E}(Y_i(t)|X_i(t),U_i,T_i>t) = \hat{E}(Y_i(t)|X_i(t),U_i) \nonumber\\= X'_i(t)\hat{\beta} + Z'_i(t)\hat{U}_i \nonumber\\
    \hat{S}(t|B_i(t),U_i) = \exp\{-\int_0^t\hat{h}_0(z)\exp(B'_i(z)\hat{\gamma}_B + \nonumber\\\hat{U}'_i \hat{\gamma}_U)dz\}
\end{eqnarray}
In linear mixed models, REs are commonly predicted targeting Best Linear Unbiased Predictors (BLUPs), \( E(U_i \mid Y_i) \), which minimize the mean squared prediction error \( E\!\left((\tilde{U} - U)^2\right) \) \citep{McCulloch2008_GLMixed}. Joint models naturally extend this to target the conditional expectation of the REs given all observed data, which now include both \( Y \) and \((D, \delta) \), hence
\[
U_{i,\mathrm{BLUP}} = E(U_i \mid Y_i, D_i, \delta_i)
\]
As they correspond to the Bayesian posterior conditional mean RE, BLUPs are also called Empirical Bayes estimates \citep{Liu2021EBtrait}.

With normal REs, BLUPs exhibit shrinkage toward the population mean, as illustrated in \citet{Liu2021EBtrait}. Likewise in joint modeling, reliance on the mean-zero prior induces shrinkage, especially with fewer longitudinal measurements. Because longitudinal and survival outcomes both depend on these effects, shrinkage can produce overly optimistic predictions, particularly when low RE values indicate poorer quality of life and survival.

\subsection{Weighted random effect predictors (WRaPs)}
For linear mixed models, \citet{mcculloch2021} demonstrated that BLUPs are not optimal when interest lies in extreme REs. Instead, they minimize a weighted mean squared error of the form
\[
E\!\left((\tilde{U}_W - U)' W(U;\lambda)(\tilde{U}_W - U)\right),
\]
with tuning parameter \(\lambda\) and \( W(U;\lambda) \) a well-chosen positive semidefinite weight function for all \(U\), with positive definite conditional expectation. Weighting generally increases overall MSE, but improves prediction in specific regions of the random effect distribution. This minimization problem is solved by 
\begin{eqnarray*}
\tilde{U}_W = \{E(W(U;\lambda) \mid Y)\}^{-1} E(W(U;\lambda)U \mid Y)
\end{eqnarray*}
We extend this to the joint modeling framework, defining the weighted random effect predictors (WRaPs) as
\begin{eqnarray}\label{weigthed_RE}
    \tilde{U}_{\mathrm{WRaP}}= \{E(W(U;\lambda)|Y,D,\delta)\}^{-1}E(W(U;\lambda)U|Y,D,\delta)
\end{eqnarray}
In the original paper, for simplicity, the authors developed a linear mixed model with a single random intercept and weight functions emphasizing errors on extreme random effect values:
\begin{eqnarray}\label{weighting_functions}
        W(u;\lambda)_{SQ} = \exp(\lambda u^2), 
        W(u;\lambda)_{AB} = \exp(\lambda |u|), \nonumber\\
        W(u;\lambda)_{CT} = I(|u| > \lambda), 
        W(u;\lambda)_{AS} = \exp(-\lambda u)
    \end{eqnarray}
We start from a joint model with random intercept and log-normal survival:
\begin{eqnarray}\label{first_model}
    Y_{ij} &=& \beta_0 + \beta_1t_{ij} + u_i + \epsilon_{ij} \nonumber\\
    \log(T_i) &\sim& N(\mu_{T,i},\sigma_T^2) \nonumber\\
    \mu_{T,i}&=&\gamma_0 + \gamma_1B_i + \gamma_2u_i
\end{eqnarray}
For ease of computation, we first predict standardized REs, \(z_i = u_i / \sigma_u\), using the WRaPs defined in (\ref{weigthed_RE}), and then recover \(\hat{u}_i\) by multiplying by the standard deviation \(\sigma_u\) \citep{mcculloch2021}. In practice, \(\sigma_u\) is first estimated from the joint model and then plugged-in for standardization. WRaPs for \(z_i\) in (\ref{first_model}) can be written as:
\begin{eqnarray} \label{new_pred censoring}
    &&z_{\mathrm{WRaP},i} \nonumber\\&= &\frac{E(W(z_i;\lambda)z_i|Y_i,D_i,\delta_i)}{E(W(z_i;\lambda)|Y_i,D_i,\delta_i)}  \nonumber\\
   & =& \frac{\int W(z_i;\lambda)z_i \frac{f_{Y_i|z_i}f_{D_i,\delta_i|z_i}f_{z_i}}{f_{Y_i,D_i,\delta_i}} dz_i}{\int W(z_i;\lambda) \frac{f_{Y_i|z_i}f_{D_i,\delta_i|z_i}f_{z_i}}{f_{Y_i,D_i,\delta_i}} dz_i} \nonumber\\
    &=& \frac{\int W(z_i;\lambda) z_i f_{Y_i|z_i}f_{D_i,\delta_i|z_i}f_{z_i} dz_i}{\int W(z_i;\lambda) f_{Y_i|z_i}f_{D_i,\delta_i|z_i}f_{z_i} dz_i}\nonumber \\ 
    &=&  \left\{
\begin{array}{ll}
\frac{\int W(z_i;\lambda)z_i f_{Y_i|z_i}f_{T_i|z_i}f_{z_i} dz_i}{\int W(z_i;\lambda) f_{Y_i|z_i}f_{T_i|z_i}f_{z_i} dz_i}\\\mathrm{if } \,  \delta_i = 1 \\
 \frac{\int W(z_i;\lambda) z_i \{1-\Phi(\frac{\log(D_i) - \mu_{T,i}}{\sigma_T})\} f_{Y_i|z_i} f_{z_i} dz_i}{\int W(z_i;\lambda)\{1- \Phi(\frac{\log(D_i) - \mu_{T,i} }{\sigma_T})\}f_{Y_i|z_i} f_{z_i} dz_i} \\\mathrm{if } \, \delta_i = 0
\end{array} \right.
\end{eqnarray}
Closed-form expressions of these WRaPs for various weight functions are given in Web Appendix A. 
\subsection{Extension to more complex joint models: Bayesian framework}\label{bayesian_approach}
With higher-dimensional REs or more flexible hazard models, deriving closed-form analytical expressions for the WRaPs is increasingly complex. Bayesian numerical strategies are then more feasible.
Efficient estimation of Markov chain Monte Carlo (MCMC) methods for standard joint models is implemented in the \texttt{R}-packages \texttt{JMbayes} and \texttt{JMbayes2} \citep{Rizopoulos2016JMbayes,JMbayes2}. For greater flexibility, we used the approach in \citet{Baghfalaki2024JointBUGS} via the \texttt{R}-packages \texttt{rjags} or \texttt{runjags}.

When using MCMC, posterior samples of subject-specific REs $U_i$ are available. These are transformed into samples of $W(Z_i;\lambda)Z_i$ and $W(Z_i;\lambda)$ by first dividing $U_i$ by \(\hat{\sigma}_u\). Per individual in the study sample, results are averaged over the MCMC iterations, and WRaPs are computed as the ratio of their posterior means, rescaled by \(\hat{\sigma}_u\). For higher-dimensional REs, we apply the above strategy separately to each dimension, using a dimension-specific tuning parameter \( \lambda \) and a one-dimensional weight function in (\ref{weighting_functions}).

\subsection{Flagging 'extreme patients'}
The clinical motivation for introducing WRaPs is to improve predictions of longitudinal outcomes, such as quality of life (QoL), and time-to-event outcomes, such as survival, for patients with extreme subject-specific REs. In case of severe diseases, identifying patients with particularly low survival probability or likely to survive with poor quality of life
(QoL) can inform discussions between clinicians, patients, and families about appropriate treatment strategies \citep{porter2025discussing}. Inspired by \citet{mcculloch2024}, we substitute BLUPs by WRaPs to improve this identification. While \citet{mcculloch2024} focused on flagging extreme random intercepts, we add a random slope and flag real, observed extreme outcomes (death or low QoL).  The tuning parameter \( \lambda \) controls the degree of emphasis placed on extreme REs and can be selected to achieve a desired balance between true and false positive rates, which makes the weighted predictions self-calibrated \citep{mcculloch2024}.

\section{Simulation study}
To illustrate the proposed methods, we conducted two type I simulation studies \citep{Heinze2024}. The first is based on a simple joint model with random intercept (1D REs) and log-normal survival. The second has a random intercept and slope (2D REs), combined with a proportional hazards Weibull model, repeating the set-up of the simulation study of \citet{rouanet2019}.

\subsection{Joint model (1D REs) with log-normal survival}\label{section1D}
We simulated for \(i = 1,\dots,1000\) and \(j = 1,\dots, N_i\):
\begin{eqnarray*}
Y_{ij} &=& 20 - 0.3\, t_j + u_i + \varepsilon_{ij} \nonumber\\
\log(T_i) &\sim& N(\mu_{T,i}, 0.5^2) \nonumber\\
\mu_{T,i} &=& -1 + B_i + 0.2\, u_i
\end{eqnarray*}
with \( \epsilon_{ij} \sim N(0, 0.81) \), \( u_i \sim N(0, 0.3)\), mutually independent, and \(B_i \sim \text{Bern}(0.5)\). Measurements were taken at time points \(t_j\) in \( \{0, 4, 12, 16, 20\}\) until death. Some values \( Y_{ij} \) were missing at random with response indicator \(R_{ij}\) satisfying
\begin{eqnarray}\label{missingness new pred}
&logit\bigl(P(R_{ij} = 1 \mid T_i > t_j, B_i, Y_{i(j-1)})\bigr) \nonumber\\
&= -1 + B_i + 0.2\, Y_{i(j-1)}
\end{eqnarray}
Our Bayesian data analysis fitted a correct joint model with unknown model parameters:
\begin{eqnarray*}
Y_{ij} &=& \beta_0 + \beta_1 t_{ij} + u_i + \varepsilon_{ij} \nonumber\\
\log(T_i) &\sim& N(\mu_{T,i}, \sigma_T^2) \nonumber\\
\mu_{T,i} &=& \gamma_0 + \gamma_1 B_i + \gamma_2 u_i
\end{eqnarray*}
using the \texttt{rjags} package. For details on prior specifications, MCMC implementation, and convergence diagnostics, see Web Appendix B. 

\begin{figure}
    \centering
    \includegraphics[width=\linewidth]{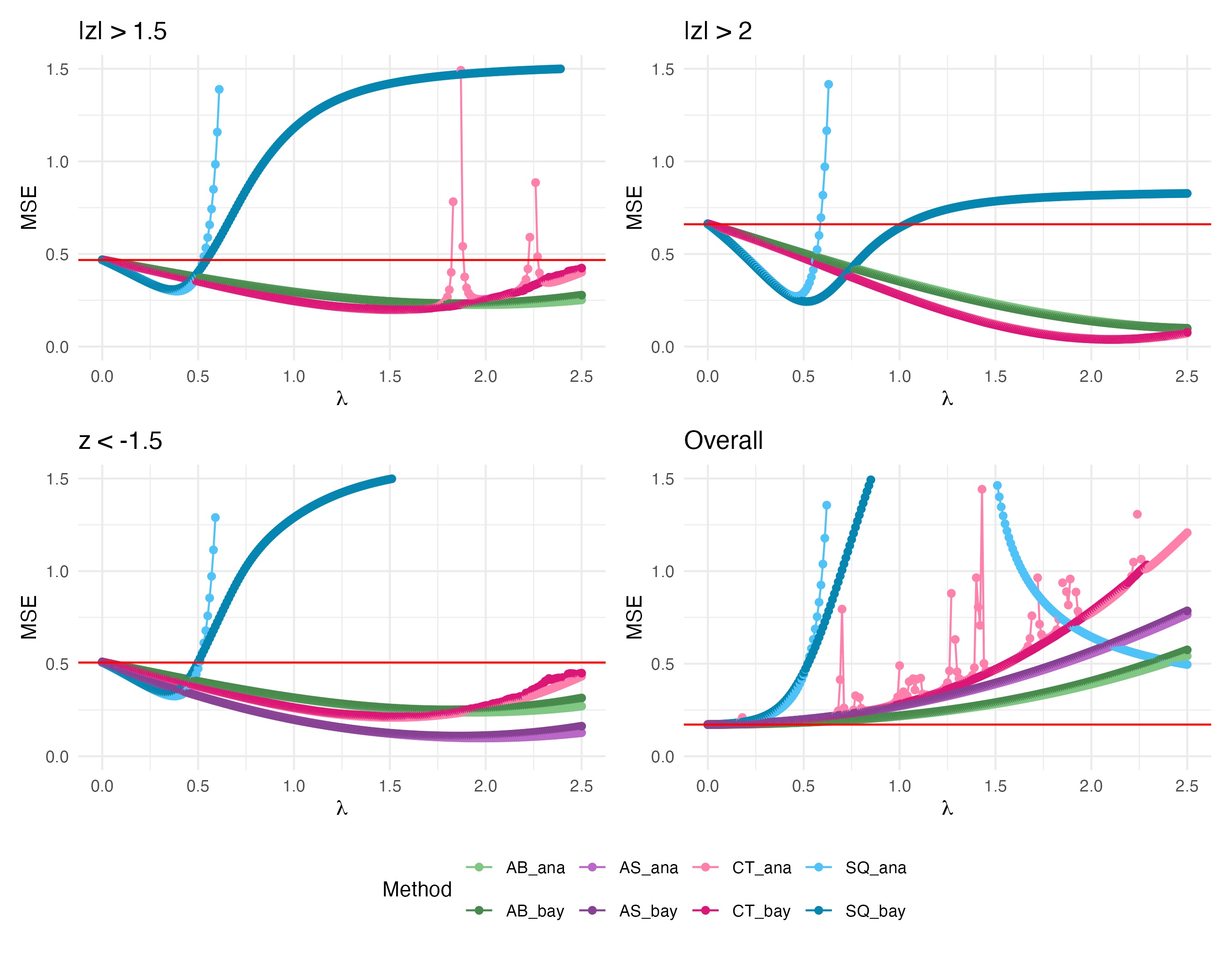}
 \caption{Mean squared error (MSE) of WRaPs as a function of the tuning parameter \(\lambda\). Panels show individuals with \(|z_i|>1.5\) (top left), \(|z_i|>2\) (top right), \(z_i<-1.5\) (bottom left) and all individuals (bottom right). Red lines indicate the MSE of the BLUPs. ``\_ana'' denotes the analytical solution, ``\_bay'' the Bayesian version.}
\label{MSE_Blups_1d}
\end{figure}

Individual WRaPs involved weight functions in (\ref{weighting_functions}) with \(\lambda = (0, 0.01, 0.02, \dots, 2.50)\) for both the analytical and the MCMC solutions. Figure~\ref{MSE_Blups_1d} shows mean squared errors (MSEs) of these predictors in subgroups \(|z_i|>1.5\), \(|z_i|>2\), \(z_i < -1.5\) and overall. For the asymmetric weight function \(AS\) we only show subgroup \(z_i< - 1.5\) results, as it aims to improve predictions for low REs. For some \(\lambda\)-values, analytical predictors \(\hat{z}_{SQ}\) and \(\hat{z}_{CT}\) have MSEs exceeding the plotting range when the denominator in (\ref{new_pred censoring}) becomes very close to zero for some individuals. The Bayesian numerical method did not suffer from this instability. 

As expected per design, Figure~\ref{MSE_Blups_1d} shows consistently lower overall MSE for BLUPs compared to WRaPs. With more extreme RE, the weighted predictions perform substantially better, demonstrating their potential when this is of particular interest. 

\subsection{Joint model (2D REs) with proportional hazards survival model}\label{2D_simulation}
In the setting with random intercept and slope, we simulated for \(i = 1,\dots,1000\) and \(j = 1,\dots, N_i\):
\begin{eqnarray*}
Y_{ij} &=& 20 - 0.3\, t_{ij} + u_{0i} + u_{1i} t_{ij} + \varepsilon_{ij} \nonumber\\
h_i(t \mid B_i, U_i) &=& 0.05 \exp(-B_i - u_{1i})
\end{eqnarray*}
with $\epsilon_{ij} \sim N(0, 0.81)$, 
$U_i \sim N\big((0,0)', \begin{pmatrix} 0.3 & -0.1 \\ -0.1 & 0.1 \end{pmatrix}\big)$, 
covariate $B \sim \text{Bern}(0.5)$, measurements at $t_j \in \{0,4,8,12,16,20\}$, 
and missingness as in (\ref{missingness new pred}). Model (\ref{Simulatie_2D}) was fitted on \(t_j\) in \(\{0,4,8,12\}\), to allow Quality of Life and survival predictions at \(t_j\) in \(\{16,20\}\).
\begin{eqnarray}\label{Simulatie_2D}
    Y_{ij} &=& \beta_0 + \beta_1t_{ij} + \beta_2X_i + \beta_3X_it_{ij} + u_{0i} + u_{1i}t_{ij} + \epsilon_{ij} \nonumber \\
    h_i(t) &=& p t^{p-1} \exp(\gamma_0 + \gamma_1 X_i + \gamma_2u_{0i} + \gamma_3u_{1i})
\end{eqnarray}
where \(X_i\) coincides with \(B_i\) in this case. Details on prior specifications, MCMC implementation, and convergence diagnostics are in Web Appendix B.

\begin{figure}[h]
  \centering
  \includegraphics[width=0.45\textwidth]{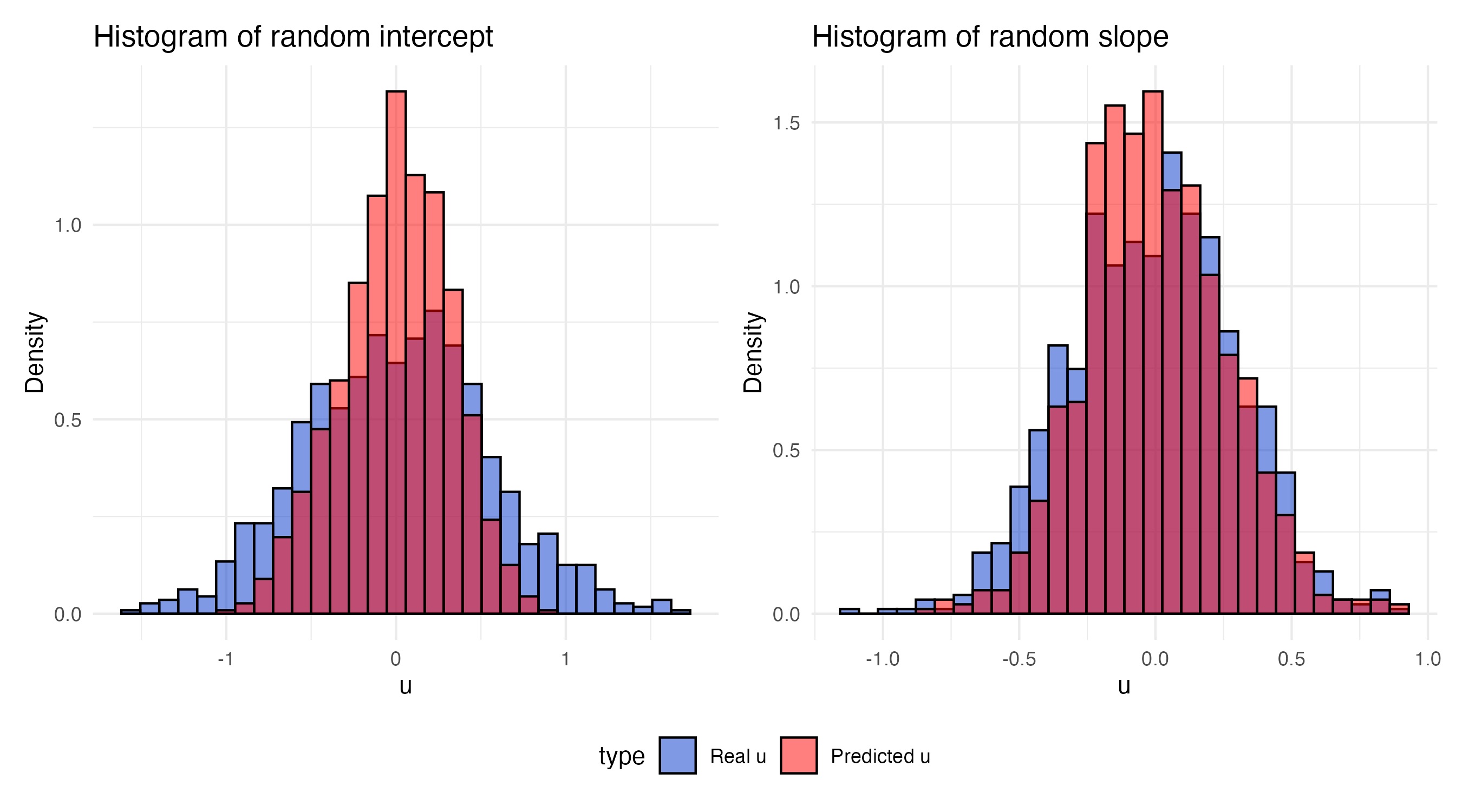}  % First figure
  \vspace{0.5cm}                                   % Vertical space between figures
  \includegraphics[width=0.45\textwidth]{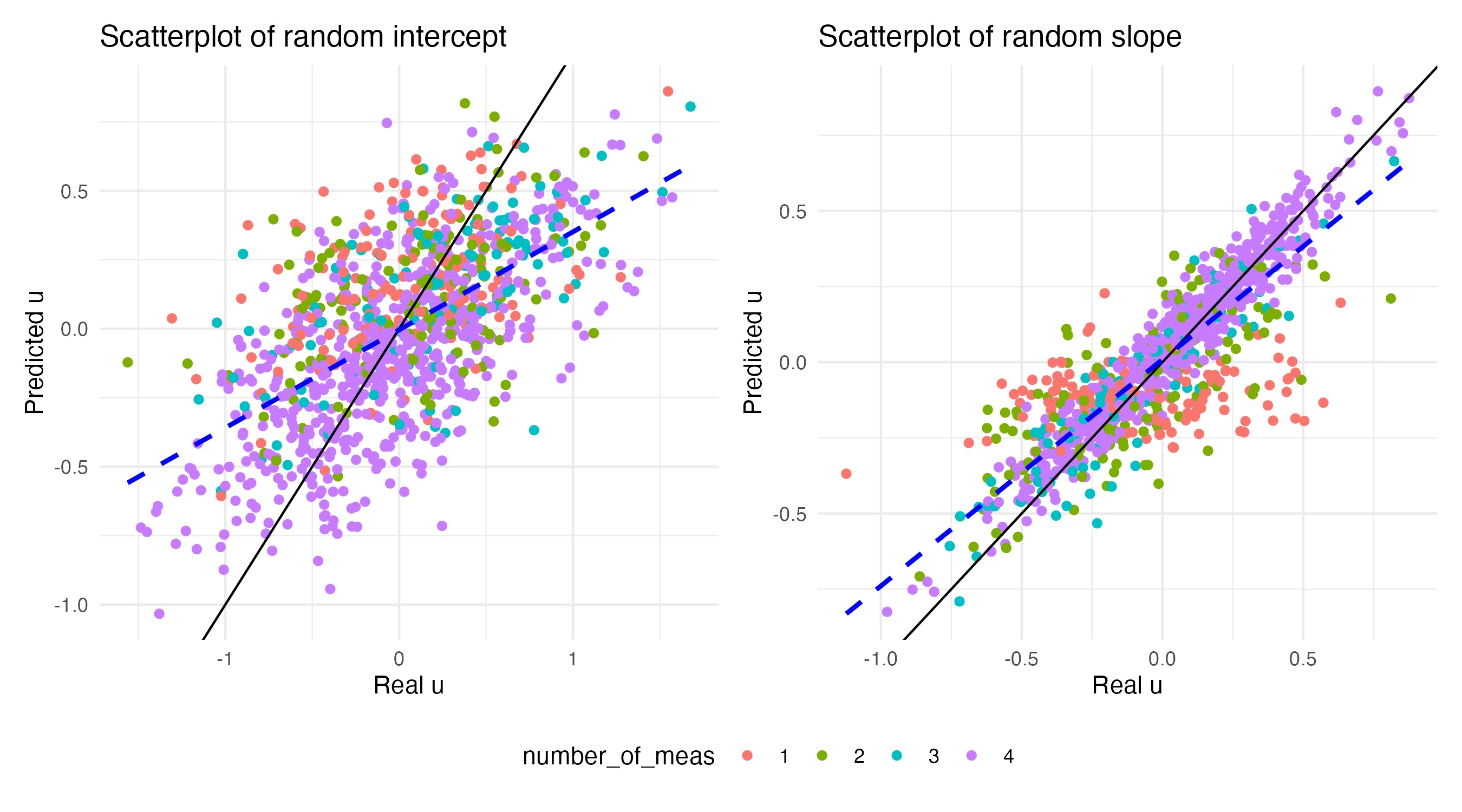}  % Second figure
  \caption{Top: Histograms of the true random effects and their corresponding BLUPs. Bottom: Scatterplots of the BLUPs versus the true random effects. The black line represents \(y=x\), and the dashed blue line shows the best-fit regression line. Points are colored according to the number of measurements for each individual.}
  \label{blups_2D}
\end{figure}

In Figure~\ref{blups_2D} histograms of the true REs and their BLUPs reveal how shrinkage is more pronounced for the random intercept than for the random slope. Scatterplots further illustrate that extreme REs are pulled toward zero. As expected, the random slope shrinks less with more measurements per individual.

With increasing RE dimension, closed-form expressions for the WRaPs are increasingly cumbersome to derive. We therefore focused on Bayesian predictions. Separate calculations for the standardized random intercept and slope used the one-dimensional weight functions in (\ref{weighting_functions}) with \(\lambda = (0, 0.02, 0.04, \dots, 3)\). For both REs, overall and subgroup-specific MSE results were similar to those obtained in section \ref{section1D}, and are omitted here.

In practice, primary interest lies in predicted future longitudinal outcomes and survival chances, not in the REs per se. For each individual in the simulated dataset, we predicted the mean outcome over future times  \(t_j\) in \(\{16, 20\}\) having fitted the joint model on data up to \(t_j = 12\). For our linear mixed model, the predicted mean outcome over future time points corresponds to the predicted mean at the midpoint. Predictions plugged in either BLUPs or WRaPs and estimated model parameters.
\begin{eqnarray}\label{prediction_qol}
    \overline{\hat{Y}}_{16-20} &=& \hat{E}(Y_{ij}|X_i,\hat{U}_i,\hat{\beta},t=\frac{16+20}{2}) \nonumber\\&=& \hat{\beta}_0 + \hat{\beta}_1\frac{16+20}{2} + \hat{\beta}_2X_i + \hat{\beta}_3 \frac{16+20}{2}X_i \nonumber \\&+& \hat{u}_{0i} + \hat{u}_{1i}\frac{16+20}{2}
\end{eqnarray}

\begin{figure}[h]
    \centering
    \includegraphics[width=0.85\linewidth]{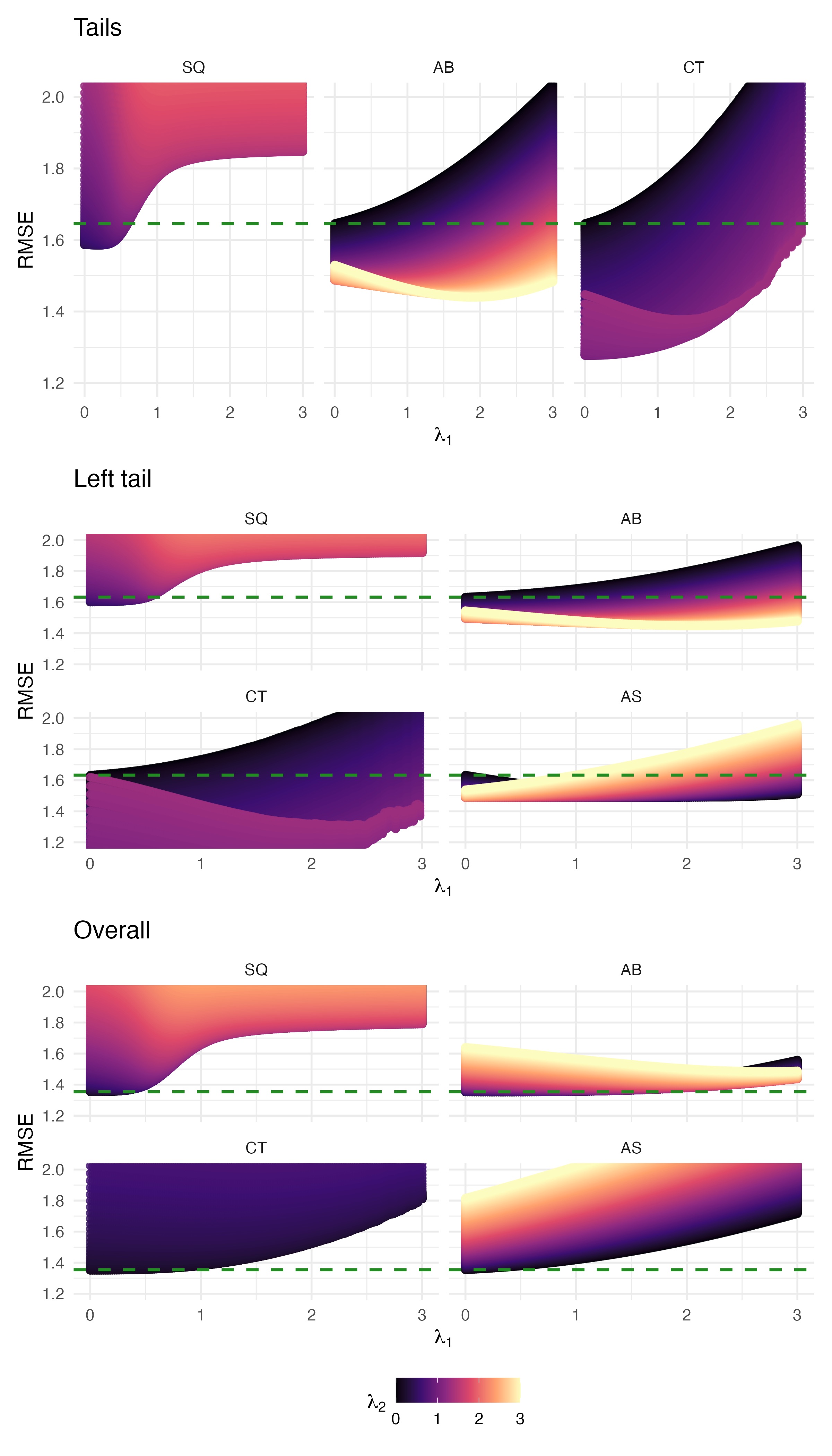}
    \caption{Root mean squared error (RMSE) of the average longitudinal outcome predictions, comparing \(\overline{\hat{Y}}_{16-20}\), with corresponding \(\overline{Y}_{16-20}\), as a function of the intercept tuning parameter \(\lambda_1\). The color of the points indicates the slope tuning parameter \(\lambda_2\). The dashed line is the BLUP reference. Upper: RMSE for subgroup with \(\overline{Y}_{16-20}\) below \(q_{0.1}\) or above \(q_{0.9}\) (“tails”). Middle: RMSE for the subgroup with \(\overline{Y}_{16-20}\) below 10th percentile (“left tail”). Bottom: RMSE over all individuals.}
    \label{MSE_qol}
\end{figure}
Figure~\ref{MSE_qol} displays the root mean squared error (RMSE) of the longitudinal outcome predictions, comparing the predicted mean outcome, \(\overline{\hat{Y}}_{16-20}\), with the true mean outcomes \(\overline{Y}_{16-20}\) in the simulated dataset. For each individual, \(\overline{Y}_{16-20}\) was calculated if at least one measurement was available within this interval. With just a single measurement observed (due to dropout or death between \(t = 16\) and \(t = 20\)) that value was used for validation. The figure reports the MSE overall, and within subgroups defined by the true mean outcomes: 'below the 10th percentile or above the 90th percentile' (Tails), and 'below the 10th percentile' (Left tail). The AS weighting method again omitted the 'tails' subgroup plot. BLUPs continued to yield the lowest overall MSE, but WRaPs substantially reduced the MSE in extreme outcome subgroups.

To illustrate WRaPs' effectiveness in identifying high-risk patients, cases were defined as those alive at \(t_j = 12\) who died before or at \(t_j = 20\), or had \(\overline{Y}_{16-20}\) below the 15th percentile of the observed mean outcomes, \(q_{0.15}\). We calculated plug-in survival estimates at \(t_j = 12\), \(\hat{S}_i(20|T_i>12)\), using BLUPs or WRaPs as for the longitudinal outcome predictions in (\ref{prediction_qol}). Patients were classified as a flag if \(\overline{\hat{Y}}_{16-20} < q_{0.15}\) or \(\hat{S}_i(20|T_i>12) <c\). For \(c\), the values \(0.1, 0.2, 0.3, 0.4\) and \(0.5\) were considered. Similarly for the 5th percentile, \(q_{0.05}\), as threshold for \(\overline{\hat{Y}}_{16-20}\).

\begin{figure}[h]
    \centering
    \includegraphics[width=\linewidth]{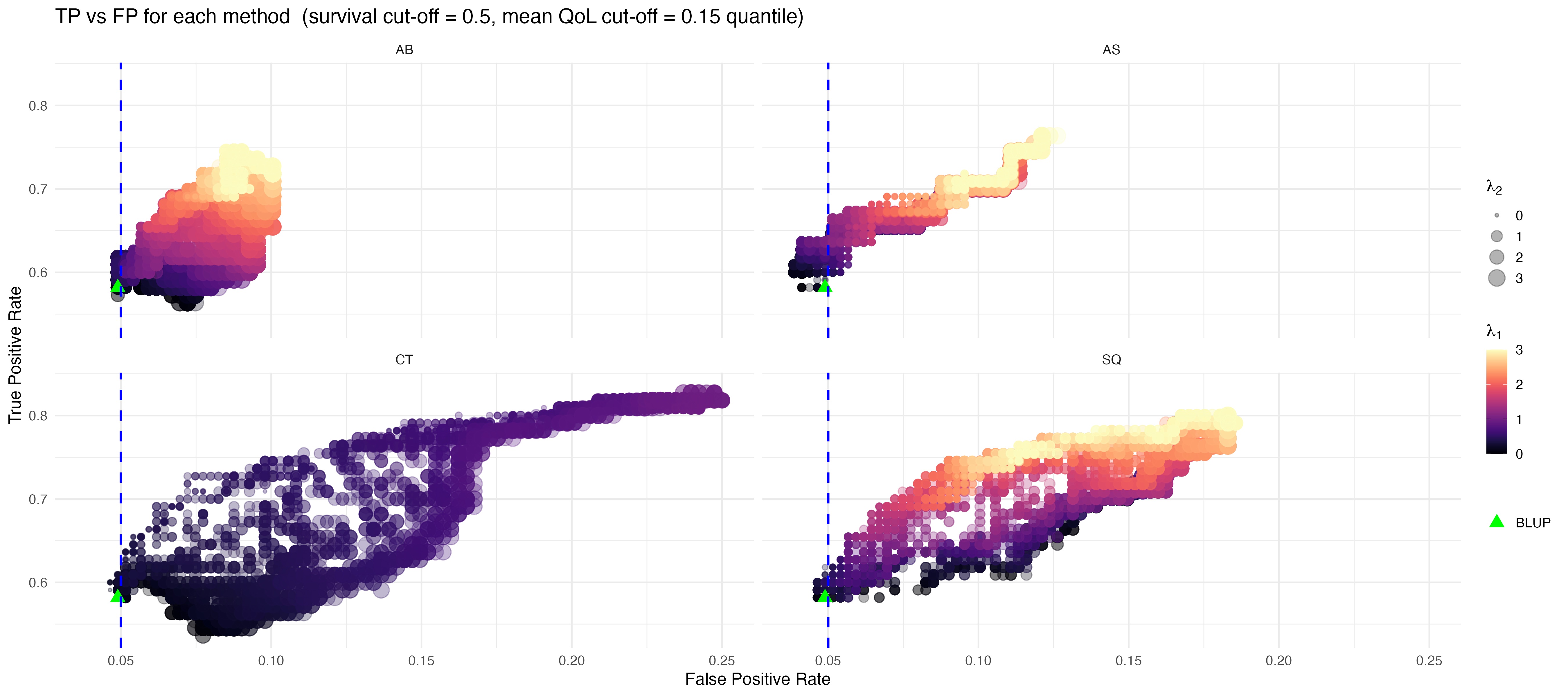}
    \includegraphics[width=\linewidth]{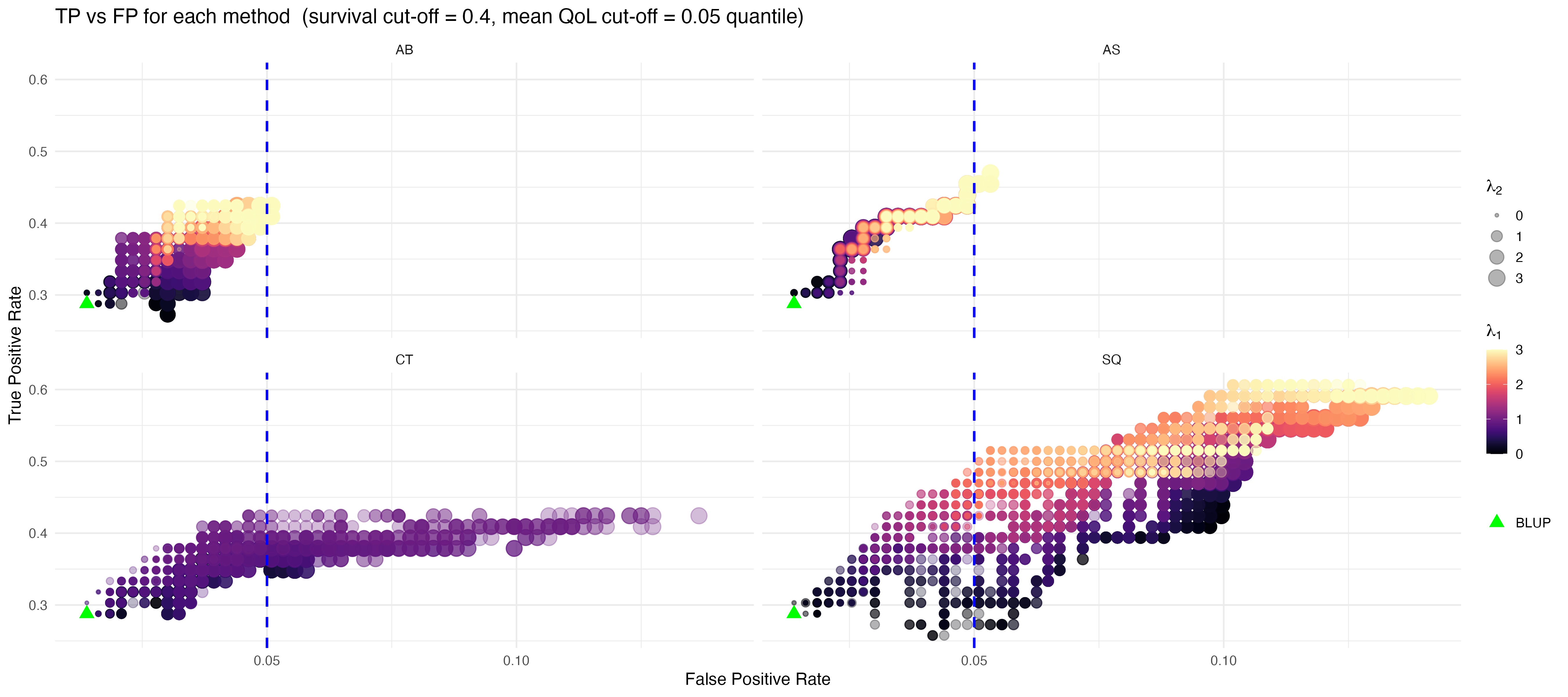}
    \caption{True Positive Rate (TPR) and False Positive Rate (FPR) for identifying individuals who, between time points \(t = (16, 20)\), either die or have a \(\overline{Y}_{16-20}\) below the \(q\)th percentile of the observed mean longitudinal outcomes in that interval. An individual was called a 'flag' if \(\hat{S}_i(20|T_i>12) <c\) or \(\overline{\hat{Y}}_{16-20} < q_x\). Upper panel: results for \(c = 0.5\) and \(q_{0.15}\). Lower panel: results for \(c = 0.4\) and \(q_{0.05}\). The blue line shows an FPR of 5\%.}
    \label{flagging}
\end{figure}

Figure~\ref{flagging} (upper) displays true and false positive rates for each weight function with \(q_{0.15}\) and \(c = 0.5\). For BLUPs, we obtain an FPR of approximately 5\% and a TPR of 58\%. With WRaPs, the TPR can be increased to nearly 65\% while maintaining an FPR of 5\%. If a higher FPR is acceptable (10-20\%), the TPR can be further improved to 75-80\%. 

In oncology settings, the cost of missing a truly at-risk patient is typically more severe than the cost of a false positive. Early detection allows clinicians to adjust treatment strategies or initiate timely discussions with patients and families about future expectations. As noted by \citet{porter2025discussing}, physicians are often overoptimistic, with adverse consequences for subsequent quality of life and psychological well-being. From a diagnostic workflow perspective, flagged patients could be prioritized for invasive confirmatory procedures, such as biopsy, thereby reducing overall burden and costs.

Figure~\ref{flagging} (bottom) shows results for the flagging rule with \(q_{0.05}\) and \(c = 0.4\). Extreme longitudinal thresholds make BLUPs perform very conservatively, yielding a 1\% FPR with TPR dropping to about 30\%. In contrast, WRaPs substantially increase the TPR to around 50\% while keeping the FPR near 5\%. SQ methods attain the best TPRs here, exceeding 60\%, with an accompanying FPR of 10\%. Results for other values of \(c\) were similar.

\section{Case study}\label{case_study}
We illustrate our method analyzing an EORTC trial that randomized 573 newly diagnosed glioblastoma patients over radiotherapy alone (\(X = 0\)) versus radiotherapy plus temozolomide (\(X=1\)). Adding temozolomide significantly improved survival while maintaining QoL \citep{Stupp2005Radiotherapy, Taphoorn2005HRQoL}. Boxplots of observed QoL at scheduled assessment times are presented in Web Appendix C1 along with descriptives of survival and missingness.

Upon fitting joint model (\ref{model_dataset}) patient-specific predictions for average QoL and survival between week 33 and week 46 (w33-w46) were derived, based on baseline covariates, prior QoL and survival status. Next, w32 survivors with poor predictions (death or average QoL below a cutoff within w33-w46) were flagged. The covariates for the survival part in (\ref{model_dataset}) were taken from \citet{Stupp2005Radiotherapy}, for the longitudinal part from \citet{Taphoorn2005HRQoL}. For simplicity, 18 patients with incomplete covariates were not included in the analysis, and follow-up was censored at week 52. 
\begin{eqnarray} \label{model_dataset} Y_{ij} &=& \beta_0 + \beta_1t_{ij} + \beta_2X_i + \beta_3 X_it_{ij} + \beta_4\mathrm{WHO}_{1i} \nonumber\\ &+& \beta_5 \mathrm{WHO}_{2i} + u_{0i} +u_{1i}t_{ij} + \epsilon_{ij} \nonumber \\ \lambda_i(t) &=& p t^{p-1}\exp(\gamma_0 + \gamma_1 X_i +\gamma_2\mathrm{WHO}_{1i} + \gamma_3 \mathrm{WHO}_{2i} \nonumber \\&+& \gamma_4 \mathrm{Surg}_{1i}  + \gamma_5 \mathrm{Surg}_{2i} + \gamma_6 \mathrm{Sex}_i + \gamma_7 \mathrm{Corti}_i \nonumber \\&+& \gamma_8 \mathrm{MMSE}_i + \gamma_9 \mathrm{Age}_i+ r_1u_{0i} + r_2u_{1i}) \end{eqnarray}

\begin{figure}[h]
    \centering
    \includegraphics[width=\linewidth]{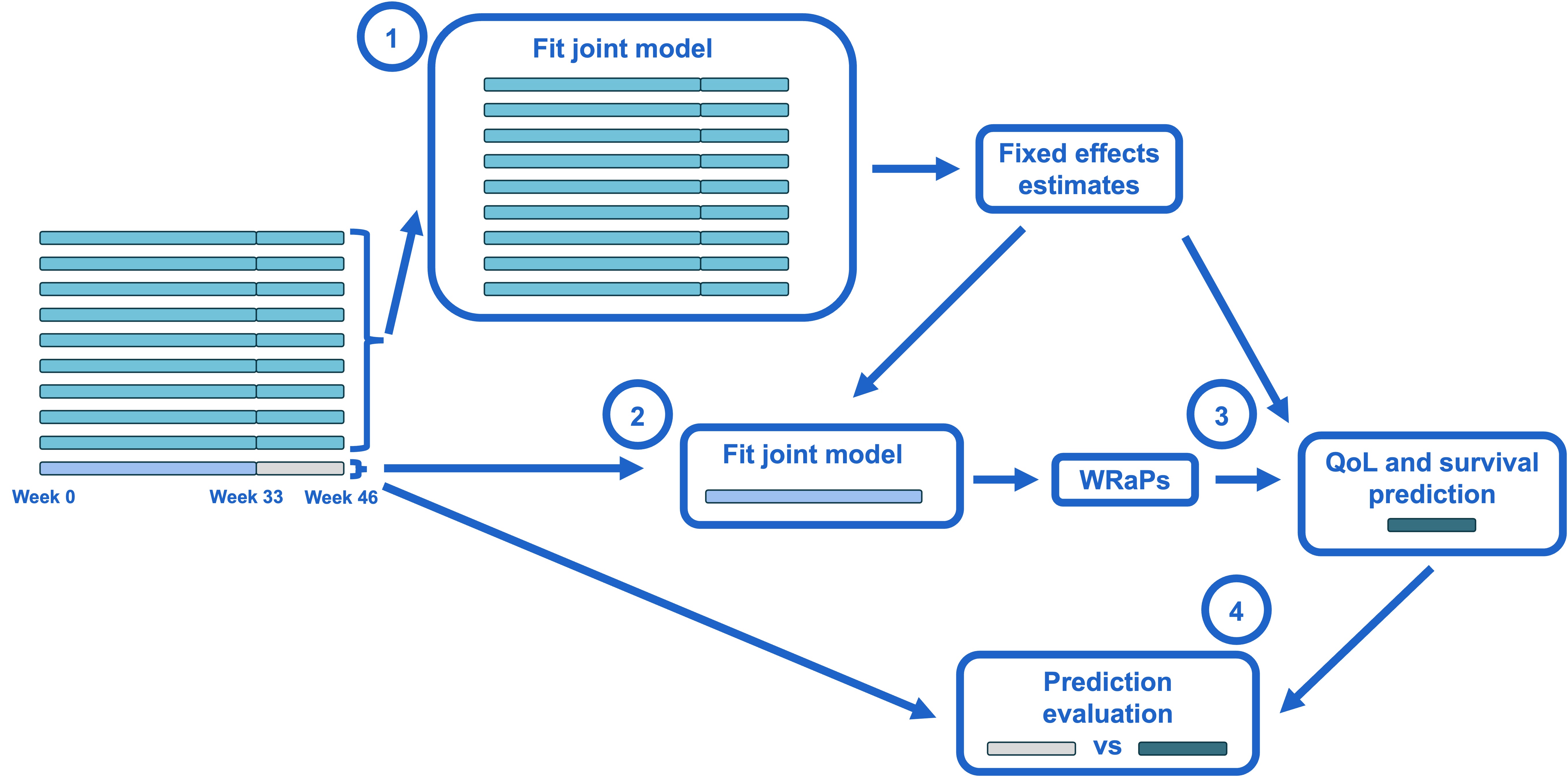}
    \caption{Prediction strategy in case study using 10-fold cross-validation. The numbers correspond with the different steps explained in Section \ref{case_study}.}
    \label{schema}
\end{figure}

Flagging performance for patients with poor outcome is assessed through 10-fold cross validation in 4 steps, see also Figure \ref{schema}: 
\begin{enumerate}
    \item Fit the joint model on the 9 folds training sets (w0-w52) with the \texttt{runjags R}-package to derive fixed effect estimates.
    \item Refit the joint model with these plug-in estimates on data up to week 32 in the left-out fold to obtain posterior samples of patient-specific REs.
    \item Derive WRaPs from the posterior RE samples of Step 2 following the Bayesian approach (Section \ref{bayesian_approach}) and involve the joint model parameter estimates from Step 1 to predict patients' (w33-w46) horizon average QoL and residual survival conditional on being alive at week 32. 
    \item Implement the flagging strategy as presented in the simulation study (Section \ref{2D_simulation}) and examine the flagging performance on the data.
\end{enumerate}

The evaluation in Step 4 was restricted to patients with available data in the week 32--46 window: 82 patients died and 131 patients reported QoL during weeks 33--46. Thus, among the 416 patients alive at week 32, only 211 were included in the analysis, as many survivors within this window did not provide QoL measurements. Moreover, among the 131 patients for whom QoL data were available, 47 provided a single measurement amounting to their observed mean QoL. Since QoL is expected to decline over time here, this approach likely overestimates the underlying interval mean. To address both issues, one could alternatively (multiple) impute missing QoL measurements given all available longitudinal information and then select extreme cases using the observed data augmented with the imputed values. Given the substantial differences between BLUP- and WRaP-based performance already seen in this simpler approach, we did not implement nor evaluate that missing data strategy here.

Here, we estimated WRaPs for all weight functions in (\ref{weighting_functions}), with \(\lambda_1, \lambda_2 = (0,0.05, 0.1,\dots, 3)\). In practice, the weight function may be prespecified in the study protocol or, as an advantage of the proposed predictors, selected a priori through simulation. Specifically, the joint model is first fitted and used to simulate outcomes over both the current and future prediction horizons. WRaPs are then computed for a range of $\lambda$-values using the simulated data to evaluate corresponding TPRs and FPRs on their future outcomes. This yields FPR--TPR curves that support clinically informed \(\lambda\) selection, which can subsequently be applied to derive WRaPs and outcome predictions for the observed data. (Code for obtaining WRaPs and corresponding FPR--TPR curves is provided in the Supplementary Material.). To enable comparison between BLUPs and WRaPs across different \(\lambda\)-choices, we considered the full grid \((0,0.05,0.1,\dots,3)\), while the proposed \(\lambda\)-selection strategy is left to future practical applications.

\begin{figure}[h]
    \centering
    \includegraphics[width=\linewidth]{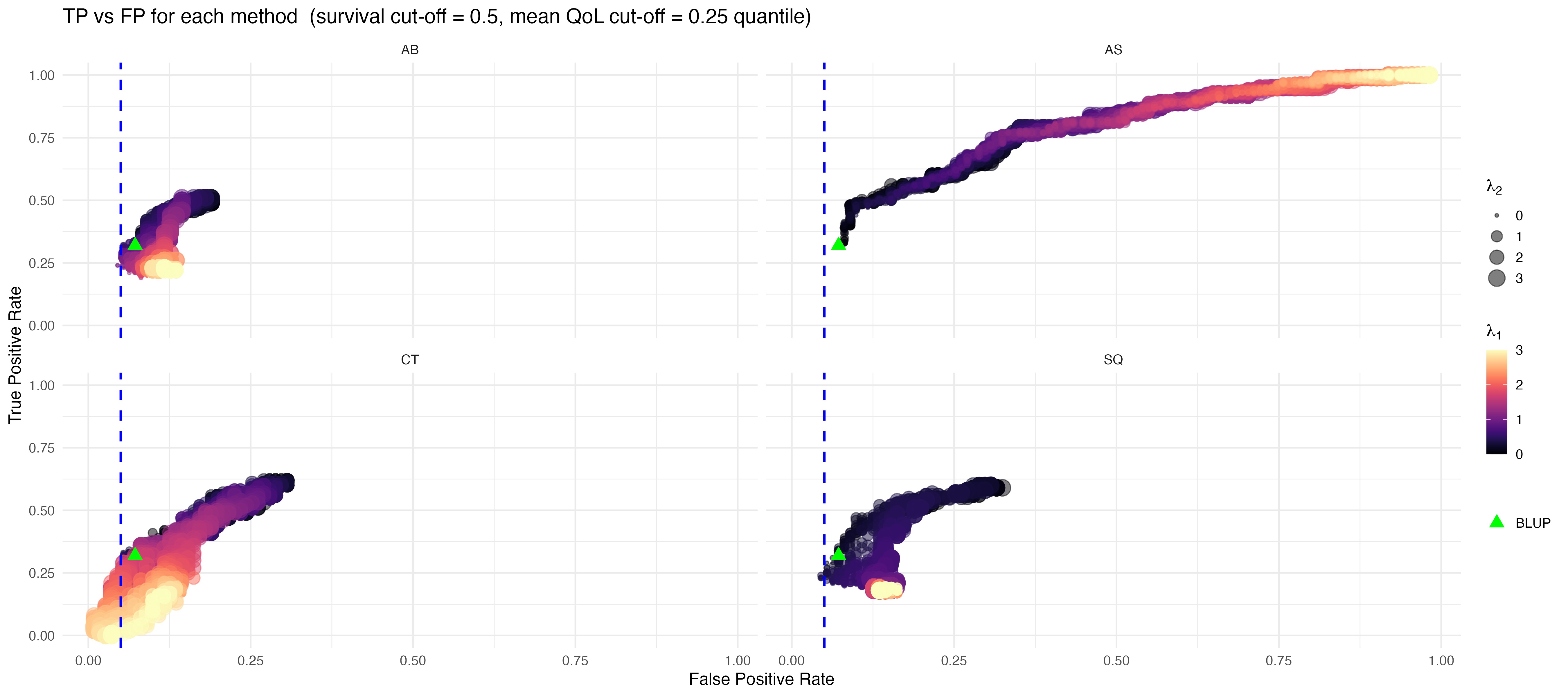}
    \includegraphics[width=\linewidth]{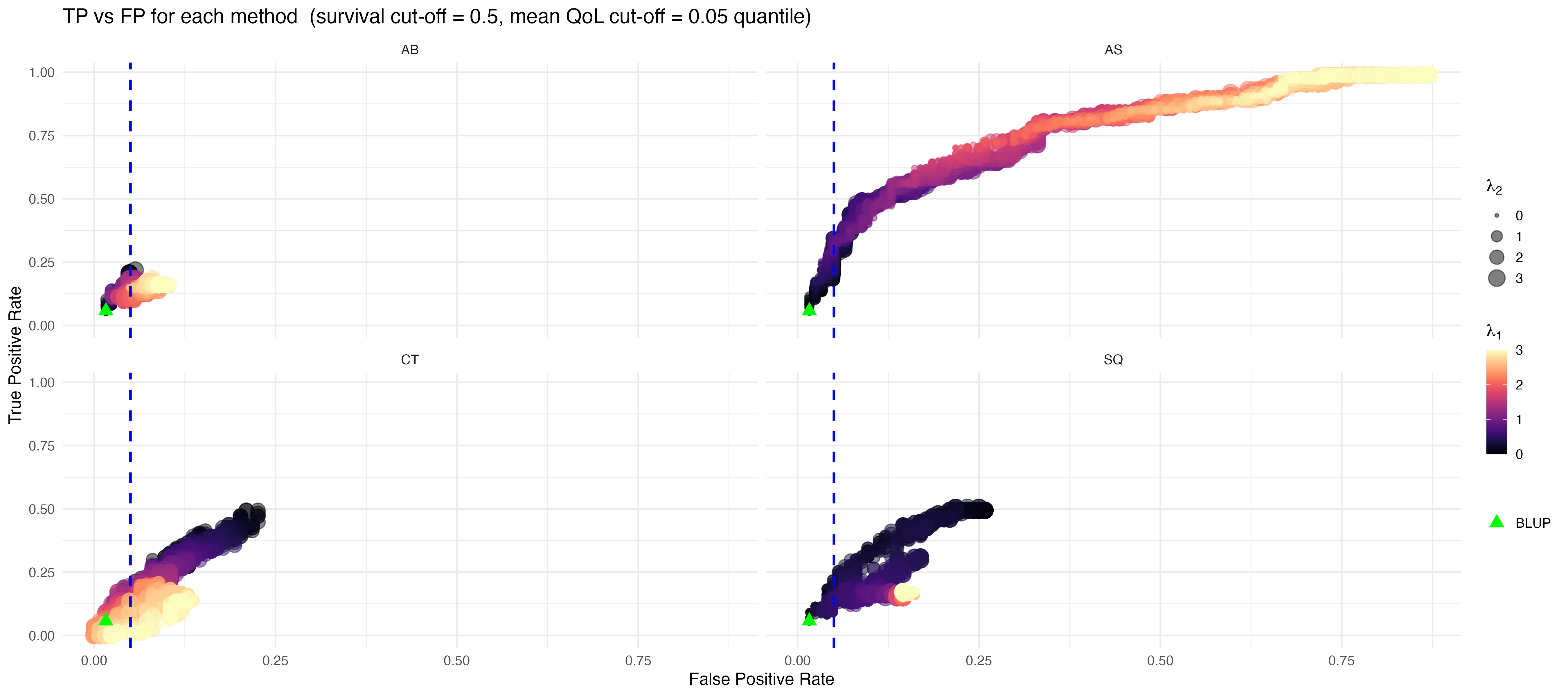}
    \caption{True Positive Rate (TPR) and False Positive Rate (FPR) for the case study, defined as in Figure~\ref{flagging}, with outcomes evaluated over $t=(33,46)$. Upper panel: $c=0.5$ with the 25th percentile QoL threshold. Lower panel: $c=0.5$ with the 5th percentile threshold. The blue line indicates an FPR of 5\%.}
    \label{flagging_data}
\end{figure}

The upper panel of Figure~\ref{flagging_data} presents results identifying cases that would die between w33-46 or had a mean QoL lower than the 25th percentile of the reported mean QoL's (mean QoL score = 50, 100 cases) with flagging rule: estimated conditional survival probability \(<\) 0.5 or predicted mean QoL lower than the 25th percentile. BLUPs had a TPR of 32\% and an FPR of 7\%, where WRaPs increased the TPR to over 50\% across all weight functions, with FPR between 10\% and 20\%. The AS method obtained a TPR of 75\% at an FPR of approximately 30\%.

We also considered the 5th percentile of the observed mean QoL threshold (mean QoL score = 33, bottom panel of Figure~\ref{flagging_data}). This resulted in 87 cases, which were poorly detected by BLUPs (TPR = 5.7\%, FPR = 1.6\%). With WRaPS, at an FPR of 5\%, the TPR exceeded 25\% for almost all weight functions, and with the AS method, a TPR of approximately 75\% was achieved at an FPR of 25\%. For this flagging rule, across all weight functions, both the positive predictive values (PPV)/negative predictive values (NPV) improve compared with the BLUPs (from 49\%/80\% with BLUPs to 67\%/83\% with WRaPs (AS weight function)). See Web Appendix C2 for plots of the PPV and NPV for both flagging rules.

\section{Discussion}
Incorporating Quality of Life (QoL) assessments into cancer and intensive care studies  moves us from the evaluation of mere survival to a joint evaluation of survival and QoL. Joint models enable this and derived individualized predictions can inform treatment decisions in which  survival benefits are weighted against the risk of deteriorating QoL. These predictions typically rely on BLUPs, which minimize the overall mean squared error of random effects, but can hinder identification of high-risk patients due to shrinkage. To address this, we propose Weighted Random effect Predictors (WRaPs) inspired by \citet{mcculloch2021}, which trade higher overall MSE for lower tail MSE, allowing us to improve identification of patients at risk of death or poor QoL within a given horizon. The proposed flagging strategy also involves tuning parameters $\lambda$ that allow to trade off true and false positive rates.

Prediction for 'extreme patients' is upgraded by focusing on subject-specific mean QoL over a well-chosen time window. This presents a more stable patient condition and limits the impact of residual variability. \citet{mcculloch2024}, who use a linear mixed model with random intercept, can flag extreme mean outcomes by flagging extreme intercepts. Our linear model with random intercept and slope makes this less straightforward. Involving a survival model further complicates the flagging approach. Extreme random effects associated with a specified threshold in the longitudinal process do not need to align with similar thresholds in survival probabilities. A more easily manageable alternative might consider a time‑specific latent effect \(\tilde{u}_t = u_0 + u_1t\), which functions as 'a random intercept at time \(t\)'. Including this quantity in both the longitudinal and survival models provides a common scale for flagging extreme patients and incorporates any correlation between intercept and slope. Such extension, which comes with additional computational complexity for the joint model and its prediction procedures is left for future research.

Our joint model specification (\ref{JM_definition}), is restricted to time-dependent covariates fully determined by baseline information. We thus avoid prediction with time-varying covariates, which would themselves require forecasting. Beyond this, the inclusion of exogenous time-dependent covariates in both the QoL and survival model would not significantly alter the developed methods. When more general hazards conditional on QoL-history are suggested, the Bayesian implementation may be well suited, but comes with considerable modeling and computational challenges.

We focused on Bayesian estimation of joint models and derived WRaPs within this framework. Using plug-in estimates for the model parameters rather than fully propagating parameter uncertainty considerably lightens the computational burden. For the linear longitudinal model, this approach is equivalent to standard Bayesian prediction, but not for the event-based outcome. In a fully Bayesian approach, WRaPs would be derived for each posterior draw of the model parameters using numerical integration, yielding a posterior survival distribution. While possible, this approach is computationally intensive. In our type I simulation study, large sample sizes and correctly specified models resulted in well-estimated parameters with unlikely qualitative affect on the plug-in estimation. This aligns with \citet{Gelman2012BayesianPredictive}, who argue that ignoring parameter uncertainty can be informally justified when predictions are not substantially affected.

How can our development help when the goal extends beyond tail accuracy to more precise estimation over the entire range of the random-effects distribution? A direction for future research is the development of hybrid predictors that use WRaPs in specific regions but retain BLUPs elsewhere, or interpolate between both within selected windows to improve random effect estimation overall.

In conclusion, WRaPs provide a practical and effective way to enhance the detection of patients at risk of severe outcomes by reducing shrinkage in the tails of the random‑effects distribution. When combined with simulation‑based tuning, they offer a flexible and clinically interpretable complement to standard BLUP‑based prediction from joint models, supporting the ongoing shift in intensive care and oncology toward more patient‑centered care.

\section*{Acknowledgements}
We are grateful for insightful discussions with Jammbe Musoro, Saskia le Cessie, Dries Reynders, Satrajit Roychoudhury and Johan Koekkoek. We also thank the European Organization for Research and Treatment of Cancer (EORTC) for permission to use the data from EORTC study 26981 for this research. The contents of this publication and methods used are solely the responsibility of the authors and do not necessarily represent the official views of the EORTC.
Part of this work is based on the master's thesis of Eline Vanderpijpen, completed at Ghent University in June 2025.

\section*{Supplementary materials}
Supplementary materials (web appendices, tables, figures, and simulation code) are available with this paper. The 2D simulation code is also available on GitHub: https://github.com/evdrpijp/WRaPs

\section*{Data availability}
Data used for the simulation study can be generated using the code provided on Github. The data underlying the case study were provided by the European Organisation For Research and Treatment of Cancer (EORTC) under licence.  The authors are not permitted to share the data and requests for access should be directed to the EORTC (https://www.eortc.org/data-sharing/).

\nocite{*}
\bibliographystyle{chicago}
\bibliography{bibliography}

@article{Taphoorn2005HRQoL,
  author    = {Taphoorn, M. J. B. and Stupp, R. and Coens, C. and Osoba, D. and Kortmann, R. and van den Bent, M. J. and Mason, W. and Mirimanoff, R. O. and Baumert, B. G. and Eisenhauer, E. and Forsyth, P. and Bottomley, A.},
  title     = {Health-related quality of life in patients with glioblastoma: a randomised controlled trial},
  journal   = {The Lancet Oncology},
  year      = {2005},
  volume    = {6},
  number    = {12},
  pages     = {937--944},
  doi       = {10.1016/S1470-2045(05)70432-0}
}

@article{Reynders2025TwoDimensionalEstimand,
  title        = {Evaluating treatment effects on longitudinal outcomes with attrition due to death: Methods for a two-dimensional estimand with a case study in Quality of Life},
  author       = {Reynders, D. and Thomassen, D. and Roychoudhury, S. and Amdal, C. D. and Musoro, J. Z. and Sauerbrei, W. and le Cessie, S. and Goetghebeur, E.},
  year         = {2025},
  journal      = {arXiv preprint},
  eprint       = {2509.25548},
  archivePrefix= {arXiv},
  primaryClass = {stat.ME},
  url          = {https://arxiv.org/abs/2509.25548},
}

@article{AhlnerElmqvist2009HRQOL,
  author    = {Ahlner-Elmqvist, M. and Jordhøy, M. S. and Bjordal, K. and Kaasa, S. and Jannert, M.},
  title     = {Health-related quality of life during the last three months of life in patients with advanced cancer},
  journal   = {Supportive Care in Cancer},
  year      = {2009},
  volume    = {17},
  number    = {2},
  pages     = {191--198},
  doi       = {10.1007/s00520-008-0477-2}
}

@article{tomer2019,
  author    = {Tomer, A. and Nieboer, D. and Roobol, M. J. and Steyerberg, E. W. and Rizopoulos, D.},
  title     = {Personalized schedules for surveillance of low-risk prostate cancer patients},
  journal   = {Biometrics},
  year      = {2019},
  volume    = {75},
  number    = {1},
  pages     = {153--162},
  doi       = {10.1111/biom.12940},
  url       = {https://doi.org/10.1111/biom.12940}
}

@article{rizopoulos2011,
  author  = {Rizopoulos, D.},
  title   = {Dynamic Predictions and Prospective Accuracy in Joint Models for Longitudinal and Time-to-Event Data},
  journal = {Biometrics},
  year    = {2011},
  volume  = {67},
  number  = {3},
  pages   = {819--829},
  doi     = {10.1111/j.1541-0420.2010.01546.x},
  url     = {https://doi.org/10.1111/j.1541-0420.2010.01546.x}
}

@article{rizopoulos2010,
  author    = {Rizopoulos, D.},
  title     = {JM: An R Package for the Joint Modelling of Longitudinal and Time-to-Event Data},
  journal   = {Journal of Statistical Software},
  year      = {2010},
  volume    = {35},
  number    = {9},
  pages     = {1--33},
  doi       = {10.18637/jss.v035.i09},
  url       = {https://doi.org/10.18637/jss.v035.i09}
}

@article{WulfsohnTsiatis1997JointModel,
  author    = {Wulfsohn, M. S. and Tsiatis, A. A.},
  title     = {A joint model for survival and longitudinal data measured with error},
  journal   = {Biometrics},
  year      = {1997},
  volume    = {53},
  number    = {1},
  pages     = {330--339},
  doi       = {10.2307/2533118},
}

@article{henderson1975,
  author    = {Henderson, C. R.},
  title     = {Best linear unbiased estimation and prediction under a selection model},
  journal   = {Biometrics},
  year      = {1975},
  volume    = {31},
  number    = {2},
  pages     = {423--447}
}

@article{robinson1991,
  author    = {Robinson, G. K.},
  title     = {That BLUP is a Good Thing: The Estimation of Random Effects},
  journal   = {Statistical Science},
  year      = {1991},
  volume    = {6},
  number    = {1},
  pages     = {15--32}
}

@article{mcculloch2021,
  author    = {McCulloch, C. E. and Neuhaus, J. M.},
  title     = {Improving Predictions When Interest Focuses on Extreme Random Effects},
  journal   = {Journal of the American Statistical Association},
  year      = {2023},
  volume    = {118},
  number    = {541},
  pages     = {504--513},
  doi       = {10.1080/01621459.2021.1938583},
  url       = {https://doi.org/10.1080/01621459.2021.1938583}
}

@article{mcculloch2024,
  author  = {McCulloch, C. E. and Neuhaus, J. M. and Boylan, R. D.},
  title   = {Flagging unusual clusters based on linear mixed models using weighted and self-calibrated predictors},
  journal = {Biometrics},
  year    = {2024},
  volume  = {80},
  number  = {2},
  pages   = {ujae022},
  doi     = {10.1093/biomtc/ujae022},
  url     = {https://doi.org/10.1093/biomtc/ujae022}
}

@article{Taylor2005IndividualizedPredictions,
  author       = {Taylor, J. M. G. and Yu, M. and Sandler, H. M.},
  title        = {Individualized predictions of disease progression following radiation therapy for prostate cancer},
  journal      = {Journal of Clinical Oncology},
  year         = {2005},
  volume       = {23},
  number       = {4},
  pages        = {816--825},
  doi          = {10.1200/JCO.2005.12.156},
  url          = {https://ascopubs.org/doi/10.1200/JCO.2005.12.156}
}

@article{Schafer2002MissingData,
  author    = {Schafer, J. L. and Graham, J. W.},
  title     = {Missing data: Our view of the state of the art},
  journal   = {Psychological Methods},
  year      = {2002},
  volume    = {7},
  number    = {2},
  pages     = {147--177},
  doi       = {10.1037/1082-989X.7.2.147}
}

@article{Stupp2005Radiotherapy,
  author    = {Stupp, R. and Mason, W. P. and van den Bent, M. J. and Weller, M. and Fisher, B. and Taphoorn, M. J. B. and Belanger, K. and Brandes, A. A. and Marosi, C. and Bogdahn, U. and Curschmann, J. and Janzer, R. C. and Ludwin, S. K. and Gorlia, T. and Allgeier, A. and Lacombe, D. and Cairncross, J. G. and Eisenhauer, E. and Mirimanoff, R. O.},
  title     = {Radiotherapy plus concomitant and adjuvant temozolomide for glioblastoma},
  journal   = {The New England Journal of Medicine},
  year      = {2005},
  volume    = {352},
  number    = {10},
  pages     = {987--996},
  doi       = {10.1056/NEJMoa043330}
}

@article{rouanet2019,
  author    = {Rouanet, A. and Helmer, C. and Dartigues, J.-F. and Jacqmin-Gadda, H.},
  title     = {Interpretation of mixed models and marginal models with cohort attrition due to death and drop-out},
  journal   = {Statistical Methods in Medical Research},
  volume    = {28},
  number    = {2},
  pages     = {343--356},
  year      = {2019},
  doi       = {10.1177/0962280217723675},
  url       = {https://pubmed.ncbi.nlm.nih.gov/28784010/}
}

@article{Liu2021EBtrait,
  author  = {Liu, S. and Kuppens, P. and Bringmann, L.},
  title   = {On the Use of Empirical Bayes Estimates as Measures of Individual Traits},
  journal = {Assessment},
  year    = {2021},
  volume  = {28},
  number  = {3},
  pages   = {845--857},
  doi     = {10.1177/1073191119885019}
}

@article{rizopoulos2009,
  author  = {Rizopoulos, D. and Verbeke, G. and Lesaffre, E.},
  title   = {Fully exponential Laplace approximations for the joint modelling of survival and longitudinal data},
  journal = {Journal of the Royal Statistical Society: Series B (Statistical Methodology)},
  year    = {2009},
  volume  = {71},
  number  = {3},
  pages   = {637--654},
  doi     = {10.1111/j.1467-9868.2008.00699.x},
  url     = {https://doi.org/10.1111/j.1467-9868.2008.00699.x}
}

@Manual{JMbayes2,
  title        = {JMbayes2: Extended Joint Models for Longitudinal and Time-to-Event Data},
  author       = {Rizopoulos, D. and Afonso, P. M. and Papageorgiou,G.},
  year         = {2025},
  version      = {0.5-7},
  note         = {R package},
  url          = {https://cran.r-project.org/web/packages/JMbayes2/JMbayes2.pdf},
}

@article{Rizopoulos2016JMbayes,
  title   = {The {R} Package {JMbayes} for Fitting Joint Models for Longitudinal and Time-to-Event Data Using {MCMC}},
  author  = {Rizopoulos, D.},
  journal = {Journal of Statistical Software},
  volume  = {72},
  number  = {7},
  pages   = {1--46},
  year    = {2016},
  doi     = {10.18637/jss.v072.i07},
  url     = {https://doi.org/10.18637/jss.v072.i07}
}

@article{Baghfalaki2024JointBUGS,
  title        = {Joint Modeling of Longitudinal Measurements and Time-to-event Outcomes Using BUGS},
  author       = {Baghfalaki, T. and Ganjali, M. and Barbieri, A. and Hashemi, R. and Jacqmin-Gadda, H.},
    journal      = {arXiv preprint},
  year         = {2024},
  eprint       = {2403.07778},
  archivePrefix= {arXiv},
  primaryClass = {stat.ME},
  url          = {https://arxiv.org/abs/2403.07778},
  doi          = {10.48550/arXiv.2403.07778}
}

@article{Benoit2014_ICUsurvival,
  author       = {Benoit,D.D. and Soares,M. and Azoulay,E.},
  title        = {Has survival increased in cancer patients admitted to the ICU? We are not sure},
  journal      = {Intensive Care Medicine},
  year         = {2014},
  volume       = {40},
  number       = {10},
  pages        = {1576--1579},
  doi          = {10.1007/s00134-014-3480-8},
  pmid         = {25217147}
}

@article{Gelman2012BayesianPredictive,
  author  = {Gelman, A. and Hwang, J. and Vehtari, A.},
  title   = {A survey of Bayesian predictive methods for model assessment, selection and comparison},
  journal = {Statistics Surveys},
  year    = {2012},
  volume  = {6},
  pages   = {1--17},
  doi     = {10.1214/12-SS102},
  url     = {https://projecteuclid.org/journals/statistics-surveys/volume-6/issue-none/A-survey-of-Bayesian-predictive-methods-for-model-assessment-selection/10.1214/12-SS102.pdf}
}

@article{guo2004separate,
  title     = {Separate and Joint Modeling of Longitudinal and Event Time Data Using Standard Computer Packages},
  author    = {Guo, X. and Carlin, B. P.},
  journal   = {The American Statistician},
  volume    = {58},
  number    = {1},
  pages     = {16--24},
  year      = {2004},
  publisher = {Taylor \& Francis},
  doi       = {10.1198/0003130042854}
}

@article{Heinze2024,
  author    = {Heinze, G. and Boulesteix, A. and Kammer, M. and Morris, T. P. and White, I. R. and the Simulation Panel of the STRATOS initiative},
  title     = {Phases of methodological research in biostatistics—Building the evidence base for new methods},
  journal   = {Biometrical Journal},
  volume    = {66},
  pages     = {2200222},
  year      = {2024},
  doi       = {10.1002/bimj.202200222},
}

@Manual{rjags,
  title = {rjags: Bayesian Graphical Models using MCMC},
  author = {Martyn Plummer},
  year = {2025},
  note = {R package version 4-17},
  url = {https://CRAN.R-project.org/package=rjags},
  version = {2.2.2-5}
}

@Manual{runjags,
  title        = {runjags: Interface Utilities, Model Templates, Parallel Computing Methods and Additional Distributions for MCMC Models in JAGS},
  author       = {Matthew Denwood and Martyn Plummer},
  year         = {2025},
  note         = {R package version 2.2.2-5},
  url          = {https://CRAN.R-project.org/package=runjags},
}

@article{Thomassen2025_missingPROs,
  author  = {Thomassen,D. and Roychoudhury,S. and Amdal,C.D. and Reynders,D. and Musoro,J.Z. and Sauerbrei,W. and Goetghebeur,E. and le Cessie,S.},
  title   = {Handling missing values in patient‐reported outcome data in the presence of intercurrent events},
  journal = {BMC Medical Research Methodology},
  year    = {2025},
  volume  = {25},
  number  = {1},
  pages   = {56},
  doi     = {10.1186/s12874-025-02510-8},
  url     = {https://link.springer.com/article/10.1186/s12874-025-02510-8},
}

@article{rubin1976,
  author    = {Rubin, D. B.},
  title     = {Inference and Missing Data},
  journal   = {Biometrika},
  year      = {1976},
  volume    = {63},
  number    = {3},
  pages     = {581--592},
  doi       = {10.2307/2335739},
  url       = {https://doi.org/10.2307/2335739}
}

@article{Griswold2021_sharedParameterModels,
  author  = {Griswold,M. E. and Talluri,R. and Zhu,X. and Su,D. and Tingle,J. and Gottesman,R. F. and Deal,J. and Rawlings, A. M. and Mosley, T. H. and B. Windham,G. and Bandeen-Roche,K.},
  title   = {Reflection on modern methods: shared-parameter models for longitudinal studies with missing data},
  journal = {International Journal of Epidemiology},
  year    = {2021},
  volume  = {50},
  number  = {4},
  pages   = {1384--1393},
  doi     = {10.1093/ije/dyab086},
  url     = {https://academic.oup.com/ije/article/50/4/1384/6296150},
}

@Book{McCulloch2008_GLMixed,
  author    = {McCulloch, C. E. and Searle, S. R. and Neuhaus, J. M.},
  title     = {Generalized, Linear, and Mixed Models},
  edition   = {2},
  publisher = {John Wiley \& Sons},
  year      = {2008},
  address   = {Hoboken, NJ},
  isbn      = {9780470073711},
}

@article{Amdal2025,
    author = {Amdal, C. D. and Falk, R. S. and Alanya, A. and others},
    title = {SISAQOL-IMI consensus-based guidelines to design, analyse, interpret, and present patient-reported outcomes in cancer clinical trials},
    journal = {The Lancet Oncology},
    year = {2025},
    pages = {e683-e693},
     volume = {26},
     doi = {10.1016/S1470-2045(25)00520-0},
   issn = {14745488},
     publisher = {Elsevier Ltd}
}

@book{book1,
	author = {Fitzmaurice, G. M. and Laird, N.M. and Ware, J. H.},
	title = {Applied Longitudinal Analysis, Second Edition},
	year = {1962},
	isbn = {978-0-470-38027-7},
	publisher = {Wiley},
}

@article{porter2025discussing,
  title={Discussing expected long-term quality of life in the ICU: effect on experiences and outcomes of patients, family, and clinicians—a randomized clinical trial},
  author={Porter, Lucy L. and Simons, Koen S. and van der Hoeven, Johannes G. and van den Boogaard, Mark and Zegers, Marieke},
  journal={Intensive Care Medicine},
  volume={51},
  number={3},
  pages={478--489},
  year={2025},
  doi={10.1007/s00134-025-07812-5},
  url={https://pubmed.ncbi.nlm.nih.gov/39992444/}
}

@article{brown2023fear,
  title={Fear of cancer recurrence and adverse cancer treatment outcomes: predicting 2- to 5-year fear of recurrence from post-treatment symptoms and functional problems in uveal melanoma survivors},
  author={Brown, S. L. and Fisher, P. and Hope-Stone, L. and Damato, B. and Heimann, H. and Hussain, R. and Cherry, M. G.},
  journal={Journal of Cancer Survivorship},
  volume={17},
  number={1},
  pages={187--196},
  year={2023},
  doi={10.1007/s11764-021-01129-0},
  url={https://link.springer.com/article/10.1007/s11764-021-01129-0}
}

@article{beil2025understanding,
  title={Understanding patient preferences: a crucial component of caring for older adults with critical conditions},
  author={Beil, M. and Benoit, D. and Guidet, B.},
  journal={Intensive Care Medicine},
  volume={51},
  number={8},
  pages={1530--1532},
  year={2025},
  doi={10.1007/s00134-025-08016-7},
  url={https://link.springer.com/article/10.1007/s00134-025-08016-7}
}

@article{basch2015patient,
  title={Patient-Reported Outcomes in Cancer Drug Development and US Regulatory Review: Perspectives From Industry, the Food and Drug Administration, and the Patient},
  author={Basch, Ethan and Geoghegan, Cindy and Coons, Stephen J. and Gnanasakthy, Ari and Slagle, Ashley F. and Papadopoulos, Elektra J. and Kluetz, Paul G.},
  journal={JAMA Oncology},
  volume={1},
  number={3},
  pages={375--379},
  year={2015},
  doi={10.1001/jamaoncol.2015.0530},
  url={https://pubmed.ncbi.nlm.nih.gov/26181187/}
}

\clearpage
\onecolumn
\appendix
\section*{Web Appendix A: Closed-form expressions of WRaPs}
On this page we give the derived close-form expressions for the Weighted Random effect Predictors in a joint model with a single intercept and a log-normal survival model, as defined in (6) of the main manuscript.
 \begin{table}[!ht]
     \centering
     \caption{Analytical expressions of WRaPs in joint model (6). \\Here \(\sigma = \frac{1}{\sqrt{n_i\sigma_\epsilon^{-2}\sigma_u^2 + \frac{\sigma_u^2\gamma_2^2}{\sigma_T^2}+1}}, \mu = \sigma_u\sigma_\epsilon^{-2}(n_i\overline{Y-\beta_0-\beta_1t}) + \frac{\gamma_2\sigma_u\sigma^2(\log(T)-\gamma_0-\gamma_1X)}{\sigma_T^2}, \mu_L = \frac{\sigma_u(\overline{Y-\beta_0-\beta_1t})}{\sigma_u^2+\frac{\sigma_\epsilon^2}{n_i}}\) and \(\sigma_L = \frac{1}{\sqrt{n_i\sigma_\epsilon^{-2}\sigma_u^2 + 1}}\).}
     \begin{tabular}{ll}
     \hline 
         Weight &  Weighted predictions\\
         function & \\
         \hline
          \(\exp(\lambda z^2)\)& \(\hat{z}_{SQ} = \left\{
                \begin{array}{ll}
           \frac{(n\sigma_\epsilon^{-2}\sigma_u^2+1+\frac{\gamma_2^2\sigma_u^2}{\sigma_T^2})\frac{\{n_i\sigma_\epsilon^{-2}\overline{(Y - \beta_0 - \beta_1t)}\sigma_u-\frac{\gamma_2\sigma_u}{\sigma_T^2}(\log(T) - \gamma_0 - \gamma_1X)\}}{n_i\sigma_\epsilon^{-2}\sigma_u^2+1+\frac{\gamma_2^2\sigma_u^2}{\sigma_T^2}}}{-2\lambda + n\sigma_\epsilon^{-2}\sigma_u^2+1+\frac{\gamma_2^2\sigma_u^2}{\sigma_T^2}} & \text{if }\delta = 1 \\
            \mu_{m,L} - \frac{\sigma_{m,L}I_8(a_m,b_m)}{1 - I_2(a_m,b_m)} & \text{if } \delta = 0
                \end{array}
                \right.\) \\
                & \\
                & with \(\mu_{m,L} = \frac{\mu_L}{1-2\lambda\sigma_L^2}, \sigma_{m,L} = \frac{\sigma_L^2}{1-2\lambda\sigma_L^2}, a_m = \frac{-\gamma_2\sigma_u\sigma_{m,L}}{\sigma_T},b_m = \frac{\log(T)-\gamma_0 -\gamma_1X - \gamma_2\sigma_u\mu_L}{\sigma_T}\) \\
          \hline 
           \(\exp(\lambda |z|)\) &  \(\hat{z}_{AB} = \left\{
\begin{array}{ll}
\frac{\exp(-\lambda\mu + \frac{\lambda^2\sigma^2}{2})\{-\sigma\phi(\frac{-\mu}{\sigma} + \lambda\sigma) + (\mu - \lambda\sigma ) \Phi (\frac{-\mu}{\sigma} + \lambda\sigma)\}
     }{\exp(\frac{\lambda^2\sigma^2}{2})[ \exp(-\lambda\mu )\Phi(\frac{-\mu}{\sigma}+\lambda \sigma) +  \exp(\lambda\mu)\{1-\Phi(\frac{-\mu}{\sigma}-\lambda \sigma)\}]} \\
     +\frac{ \exp(\lambda\mu + \frac{\lambda^2\sigma^2}{2})[\sigma \phi(\frac{-\mu}{\sigma - \lambda \sigma})+ (\lambda \sigma + \mu) \{1 - \Phi (\frac{-\mu}{\sigma - \lambda \sigma})\}]}{\exp(\frac{\lambda^2\sigma^2}{2})[ \exp(-\lambda\mu )\Phi(\frac{-\mu}{\sigma}+\lambda \sigma) +  \exp(\lambda\mu)\{1-\Phi(\frac{-\mu}{\sigma}-\lambda \sigma)\}]}& \text{if }\delta = 1 \\
     \frac{\hat{z}_{AB}^{L,n} -\mu_L \exp(\frac{\lambda^2\sigma_L^2}{2})\{\exp(-\lambda\mu_L)I_1(a_1,b_1,\frac{-\mu_L}{\sigma_L} - \lambda\sigma_L) + \exp(\lambda\mu_L)I_3(a_2,b_2,\frac{-\mu_L}{\sigma_L} - \lambda\sigma_L)\}}{\hat{z}_{AB}^{L,d} - \exp(\frac{\lambda^2\sigma_L^2}{2})\{\exp(-\lambda\mu_L)I_1(a_1,b_1,\frac{-\mu_L}{\sigma_L} + \lambda\sigma_L) + \exp(\lambda\mu_L)I_3(a_2,b_2,\frac{-\mu_L}{\sigma_L} - \lambda\sigma_L)\}} & \\
     -\frac{2\lambda \sigma_L^2 \phi(-\frac{\mu_L}{\sigma_L})\Phi(-\frac{a\mu_L}{\sigma_L} + b)}{\hat{z}_{AB}^{L,d} - \exp(\frac{\lambda^2\sigma_L^2}{2})\{\exp(-\lambda\mu_L)I_1(a_1,b_1,\frac{-\mu_L}{\sigma_L} + \lambda\sigma_L) + \exp(\lambda\mu_L)I_3(a_2,b_2,\frac{-\mu_L}{\sigma_L} - \lambda\sigma_L)\}} &\\- \frac{\sigma_L \exp(\frac{\lambda^2\sigma_L^2}{2})}{\hat{z}_{AB}^{L,d} - \exp(\frac{\lambda^2\sigma_L^2}{2})\{\exp(-\lambda\mu_L)I_1(a_1,b_1,\frac{-\mu_L}{\sigma_L} + \lambda\sigma_L) + \exp(\lambda\mu_L)I_3(a_2,b_2,\frac{-\mu_L}{\sigma_L} - \lambda\sigma_L)\}}  &\\.[\exp(\lambda\mu_L)\{\lambda\sigma_L I_3(a_2,b_2,-\frac{\mu_L}{\sigma_L} - \lambda\sigma_L)
     + aI_6(a_2,b_2,-\frac{\mu_L}{\sigma_L} - \lambda\sigma_L)\} &\\ - \exp(-\lambda\mu_L)\{\lambda\sigma_L I_1(a_1,b_1,-\frac{\mu_L}{\sigma_L} + \lambda\sigma_L) + aI_4(a_1,b_1,-\frac{\mu_L}{\sigma_L} + \lambda\sigma_L)\}] & \text{if } \delta = 0
\end{array}
\right.  \) \\
& \\
& with \(a_1 =a_2= \frac{-\gamma_2\sigma_u\sigma_L}{\sigma_T}\), \(b_1 = -a_1\lambda\sigma_L + \frac{\log(T)-\gamma_0 -\gamma_1X - \gamma_2\sigma_u\mu_L}{\sigma_T}\), \\&\(b_2 = a_1\lambda\sigma_L + \frac{\log(T)-\gamma_0 -\gamma_1X - \gamma_2\sigma_u\mu_L}{\sigma_T}\), \\
& \(\hat{z}_{AB}^{L,n} =
\exp(-\lambda \mu_{L} + \frac{\lambda^{2}\sigma_{L}^{2}}{2})
\{
- \sigma_{L} \, \phi(-\frac{\mu_{L}}{\sigma_{L}} + \lambda \sigma_{L})
- (\lambda \sigma_{L}^{2} - \mu_{L})
\Phi(-\frac{\mu_{L}}{\sigma_{L}} + \lambda \sigma_{L})
\}
+\) \\
&\(\exp(\,\lambda \mu_{L} + \frac{\lambda^{2}\sigma_{L}^{2}}{2})
[
\sigma_{L} \, \phi(-\frac{\mu_{L}}{\sigma_{L}} - \lambda \sigma_{L})
+ (\lambda \sigma_{L}^{2} + \mu_{L})
\{1 - \Phi(-\frac{\mu_{L}}{\sigma_{L}} - \lambda \sigma_{L})\}
]\) and \\
& \(\hat{z}_{AB}^{L,d} =\exp(-\lambda \mu_{L} + \frac{\lambda^{2}\sigma_{L}^{2}}{2})
\Phi(-\frac{\mu_{L}}{\sigma_{L}} + \lambda \sigma_{L})
+ \exp(\lambda \mu_{L} + \frac{\lambda^{2}\sigma_{L}^{2}}{2})
\{
1 - \Phi(-\frac{\mu_{L}}{\sigma_{L}} - \lambda \sigma_{L})
\}\) \\
     \hline 
     \(I_{|z|>\lambda}\) & \( \hat{z}_{CT} = \left\{
\begin{array}{ll}
\frac{- \sigma \phi(\frac{-\lambda - \mu}{\sigma}) + \mu \Phi(\frac{-\lambda - \mu}{\sigma}) + \sigma \phi(\frac{\lambda - \mu}{\sigma}) + \mu \{1- \Phi(\frac{\lambda - \mu}{\sigma})\}}{\Phi(\frac{-\lambda - \mu}{\sigma}) + 1 - \Phi(\frac{\lambda - \mu}{\sigma})} & \text{if }\delta = 1 \\
\frac{\hat{z}_{CT}^{L,n} - \mu_LI_1(a,b,\frac{-\lambda-\mu_L}{\sigma_L}) -\sigma_LI_7(a,b,\frac{-\lambda-\mu_L}{\sigma_L})- \mu_L I_3(a,b,\frac{\lambda - \mu_L}{\sigma_L}) - \sigma_L I_9(a,b,\frac{\lambda - \mu_L}{\sigma_L})}{\Phi(\frac{-\lambda-\mu_L}{\sigma_L}) + 1 - \Phi(\frac{\lambda-\mu_L}{\sigma_L}) - I_1(a,b,\frac{-\lambda-\mu_L}{\sigma_L}) - I_3(a,b,\frac{\lambda-\mu_L}{\sigma_L})} & \text{if } \delta = 0
\end{array}
\right.\) \\
&\\
& with \(\hat{z}_{CT}^{L,n} = -\sigma_L\phi(\frac{-\lambda-\mu_L}{\sigma_L})+\mu_L\Phi(\frac{-\lambda-\mu_L}{\sigma_L}) + \sigma_L\phi(\frac{\lambda-\mu_L}{\sigma_L}) + \mu_L\{1-\Phi(\frac{\lambda-\mu_L}{\sigma_L})\}\)\\

     \hline 
     \(\exp(-\lambda z)\) & \( \hat{z}_{AS} = \left\{
\begin{array}{ll}
\mu - \sigma^{2}\lambda & \text{if }\delta = 1 \\
\mu_L - \frac{\sigma_L \{\sigma_L\lambda + aI_5(a_1,b_1) - \lambda \sigma_L I_2(a_1,b_1)\}}{1-I_2(a_1,b_1)} & \text{if } \delta = 0
\end{array}
\right.  \) \\
     \hline
     \end{tabular} 
     \label{table newly pred}
 \end{table}

\newpage
\section*{Web Appendix B: Bayesian analysis}
All joint models in the main manuscript were fitted in a Bayesian framework. The corresponding \texttt{BUGS} code for the model used in the two-dimensional simulation study is provided at the end of this section. Code for the remaining models is omitted, as these differ only in minor details.

\paragraph{Prior distributions}
For both simulation studies, the fixed effects in the longitudinal and survival models were assigned independent normal priors with mean zero and precision 0.001, corresponding to a standard deviation of 31.6. The residual variance of the longitudinal model, $\epsilon_{ij}$, was assigned an inverse-Gamma prior with shape and scale parameters equal to 0.01, reflecting weakly informative prior knowledge. The covariance matrix of the random effects was assigned an inverse-Wishart prior with 3 degrees of freedom and the identity matrix as scale matrix. In the one-dimensional simulation study, inverse-Gamma priors with shape and scale equal to 0.01 were used for the variance of the random effect and for the variance of the survival times. In the two-dimensional simulation study, the shape parameter of the Weibull survival distribution was assigned a $\text{Gamma}(1,1)$ prior. These prior choices were motivated by \citet{Baghfalaki2024JointBUGS}. 

For the case study, the same prior specifications as in the two-dimensional simulation were used, with two modifications to improve convergence: the Weibull shape parameter was assigned a $\text{Gamma}(8,5)$ prior, and the precision of the normal priors for the survival coefficients associated with the random effects was set to 0.01, yielding a more informative prior.

\paragraph{Computation and MCMC settings}
Models were implemented in \texttt{R} using the \texttt{rjags} package, which performs posterior sampling via Markov chain Monte Carlo (MCMC). For the case study, the \texttt{runjags} package was used to allow parallel sampling of multiple chains. In all analyses, three parallel chains were run.

Initial values for the longitudinal intercept $\beta_0$ were set to the observed mean outcome at baseline (time zero) in the untreated group. In the simulation studies, initial values for censored event times were set to the censoring time plus 2, 5, and 10 for the three chains, respectively. In the case study, these offsets were set to 1, 2, and 3. Initial values for the fixed effects in the longitudinal model were drawn from a standard normal distribution. For the survival model, a Weibull model without random effects was first fitted, and the resulting parameter estimates were used as initial values for the fixed effects and the Weibull shape parameter.

Models were compiled using the \texttt{jags.model} function. In the simulation studies, a burn-in of 5{,}000 iterations was used, after which posterior samples were drawn using the \texttt{coda.samples} function, yielding 100{,}000 samples across the three chains with a thinning interval of 10. Posterior means were used as parameter estimates. For the case study, an adaptation phase of 10{,}000 iterations and a burn-in of 500{,}000 iterations were used, followed by 10{,}000 posterior samples with thinning of 10. Parameter estimates for the longitudinal model and the random-effects distribution were obtained as posterior means, while posterior medians were used for the survival model.

\paragraph{Convergence diagnostics}
Convergence and sampling efficiency were assessed using the Gelman--Rubin diagnostic, trace plots, and effective sample size (ESS) for all model parameters.

\begin{figure}[h]
    \centering
    \includegraphics[width=0.6\linewidth]{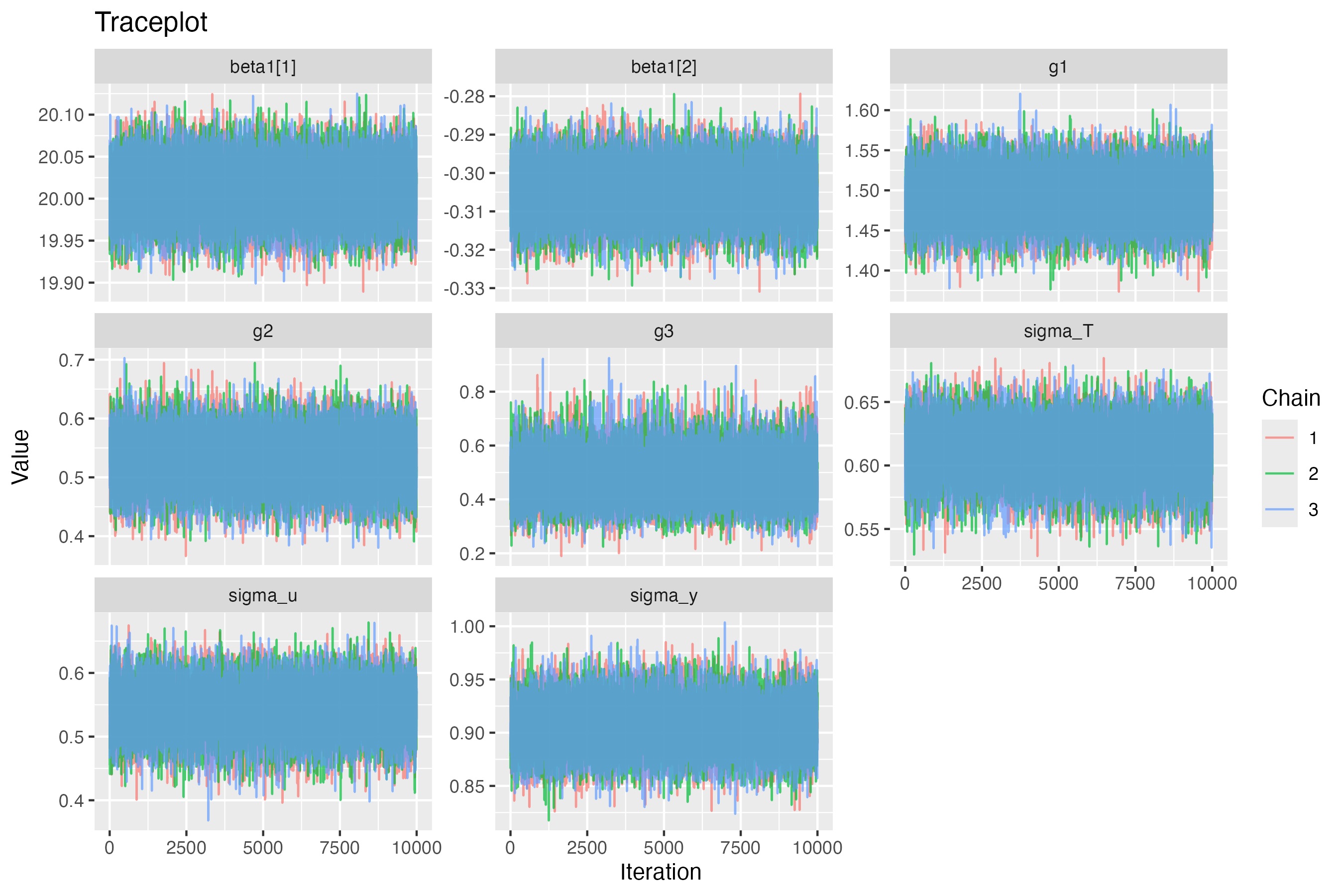}
    \caption{Traceplot of parameters in 1D joint model of the simulation study.}
    \label{trace_sim1}
\end{figure}

\begin{figure}[h]
    \centering
    \includegraphics[width=0.7\linewidth]{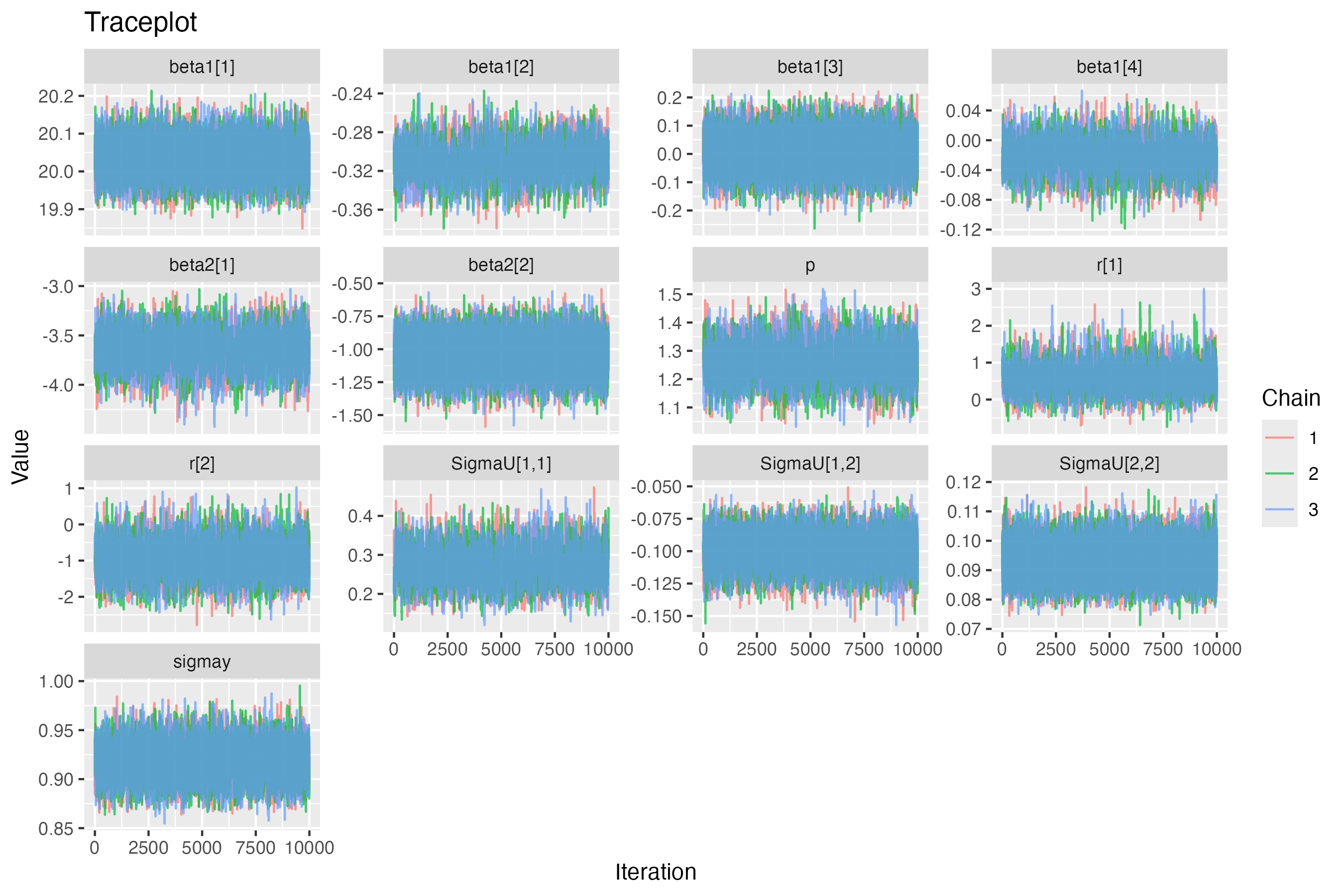}
    \caption{Traceplot of parameters in 2D joint model of the simulation study.}
    \label{trace_sim2}
\end{figure}

In the simulation study, convergence was adequate. The effective sample size (ESS) for all model parameters ranged from 11861 to 30000 in the one-dimensional random effects setting and from 1261 to 10468 in the two-dimensional setting. The Gelman-Rubin statistics had a maximum value of 1.01, further indicating adequate convergence. This was also supported by the trace plots, which exhibited good mixing and high density, as shown in Figures \ref{trace_sim1} and \ref{trace_sim2}.


\begin{figure}
    \centering
    \includegraphics[width=0.45\linewidth]{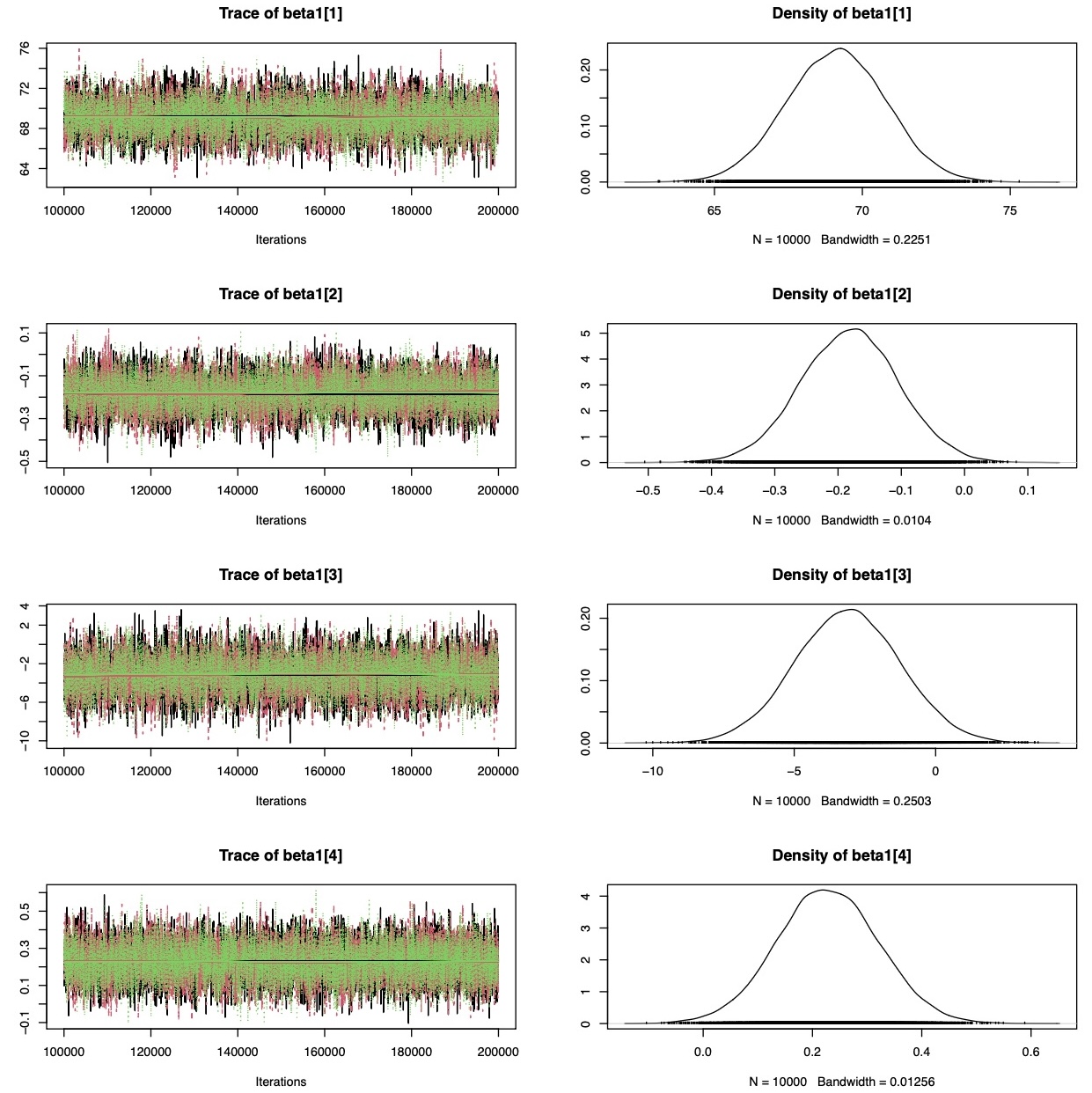}
     \includegraphics[width=0.45\linewidth]{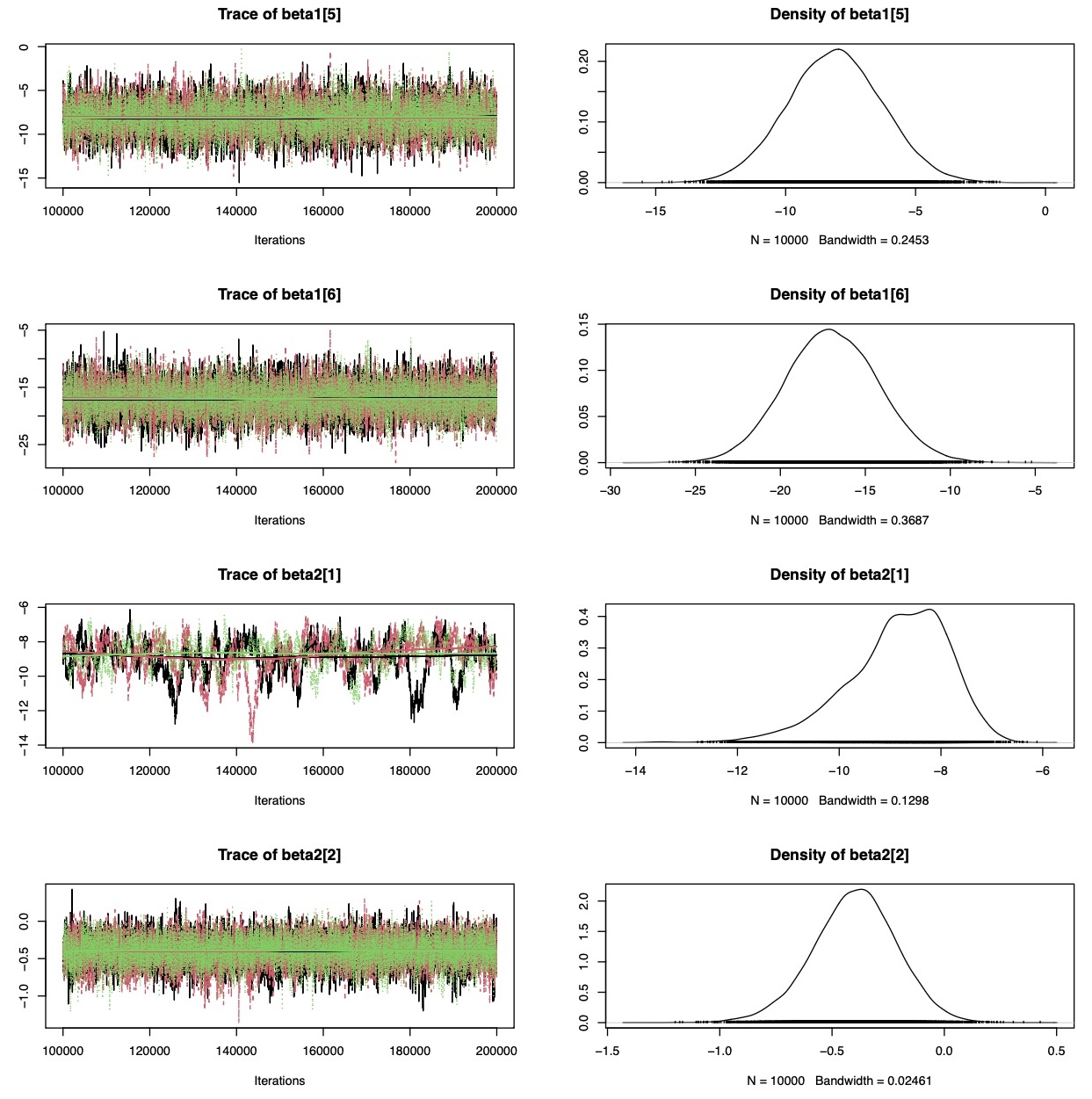}
      \includegraphics[width=0.45\linewidth]{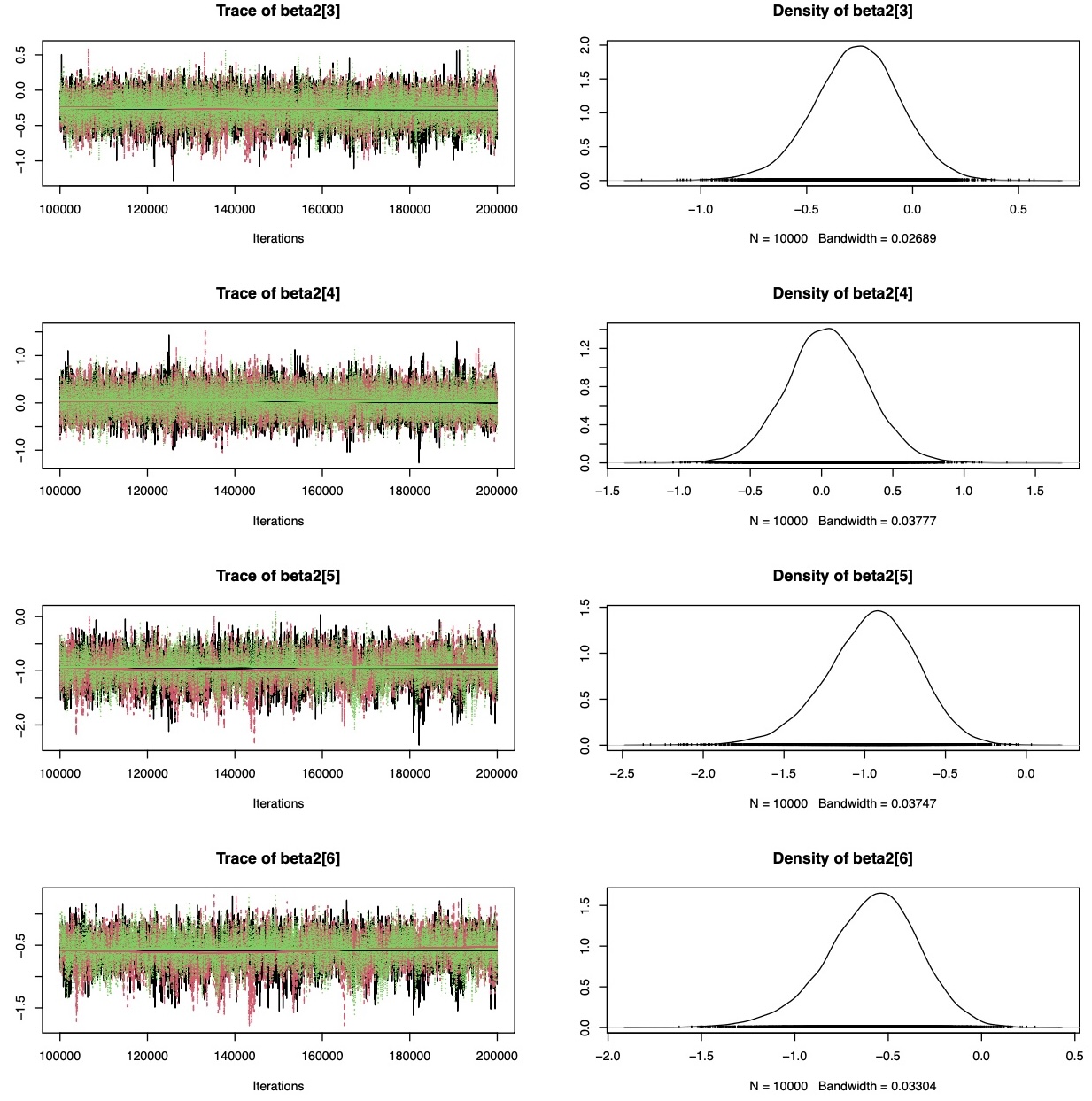}
       \includegraphics[width=0.45\linewidth]{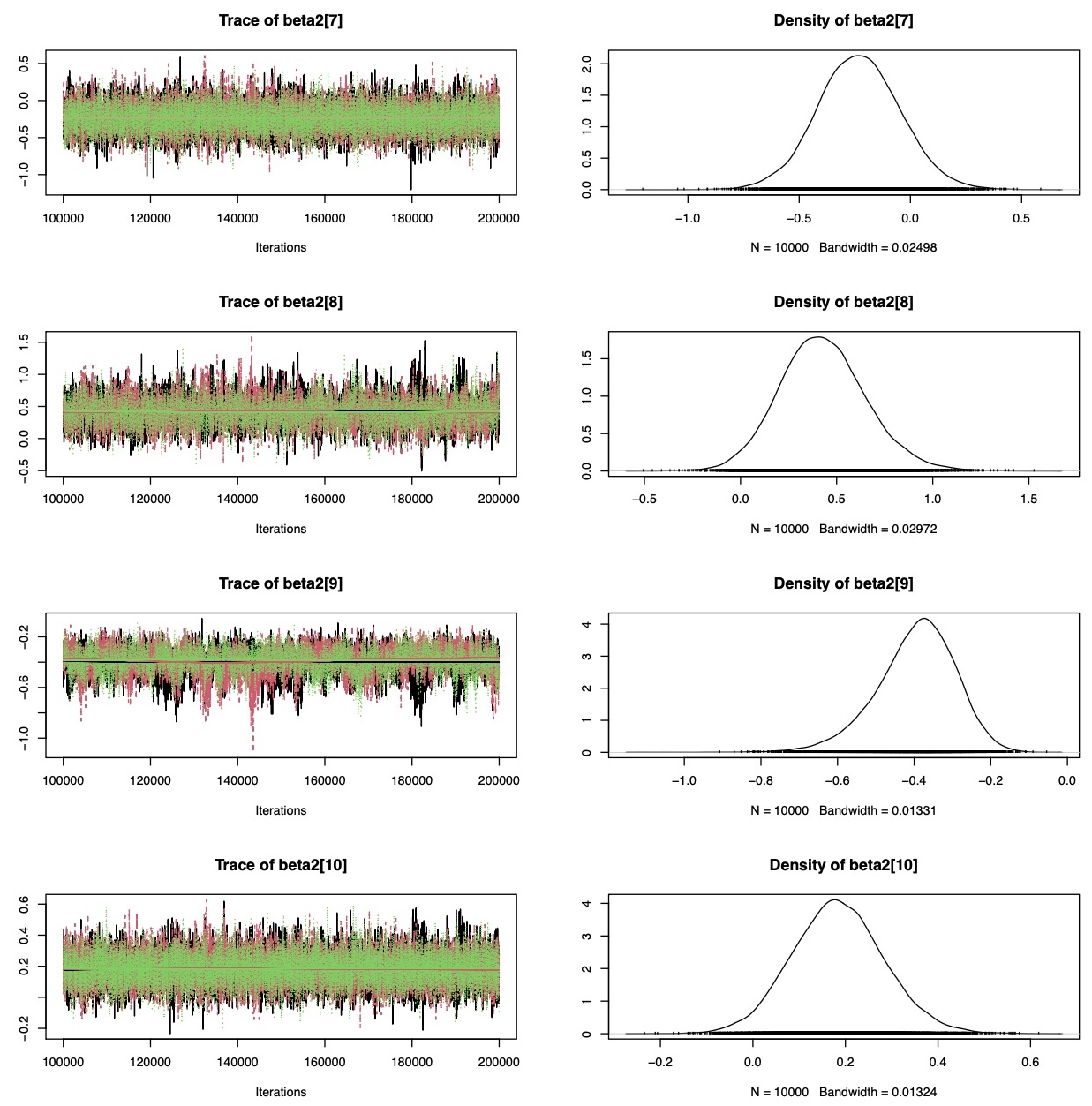}
        \includegraphics[width=0.45\linewidth]{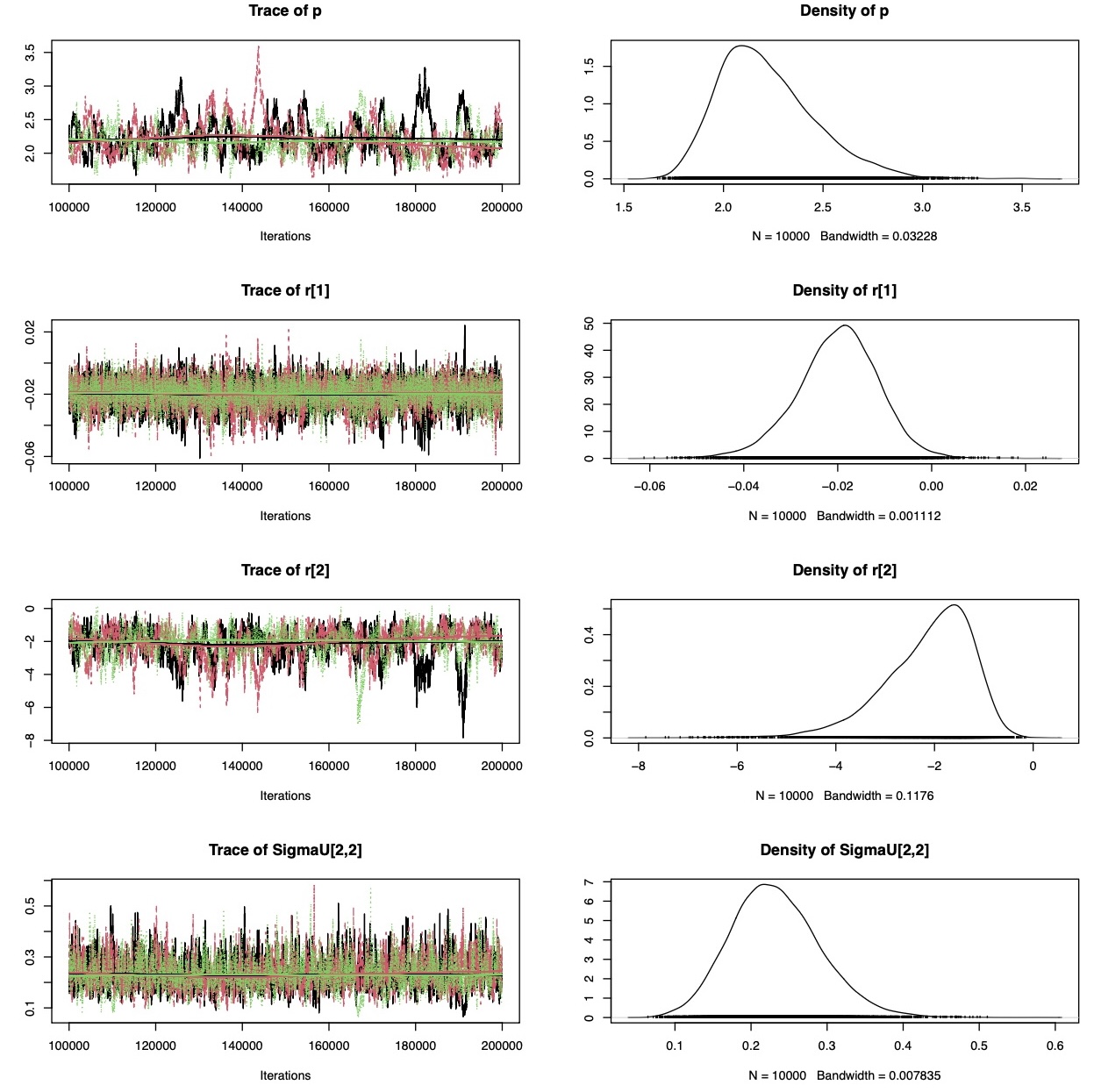}
         \includegraphics[width=0.45\linewidth]{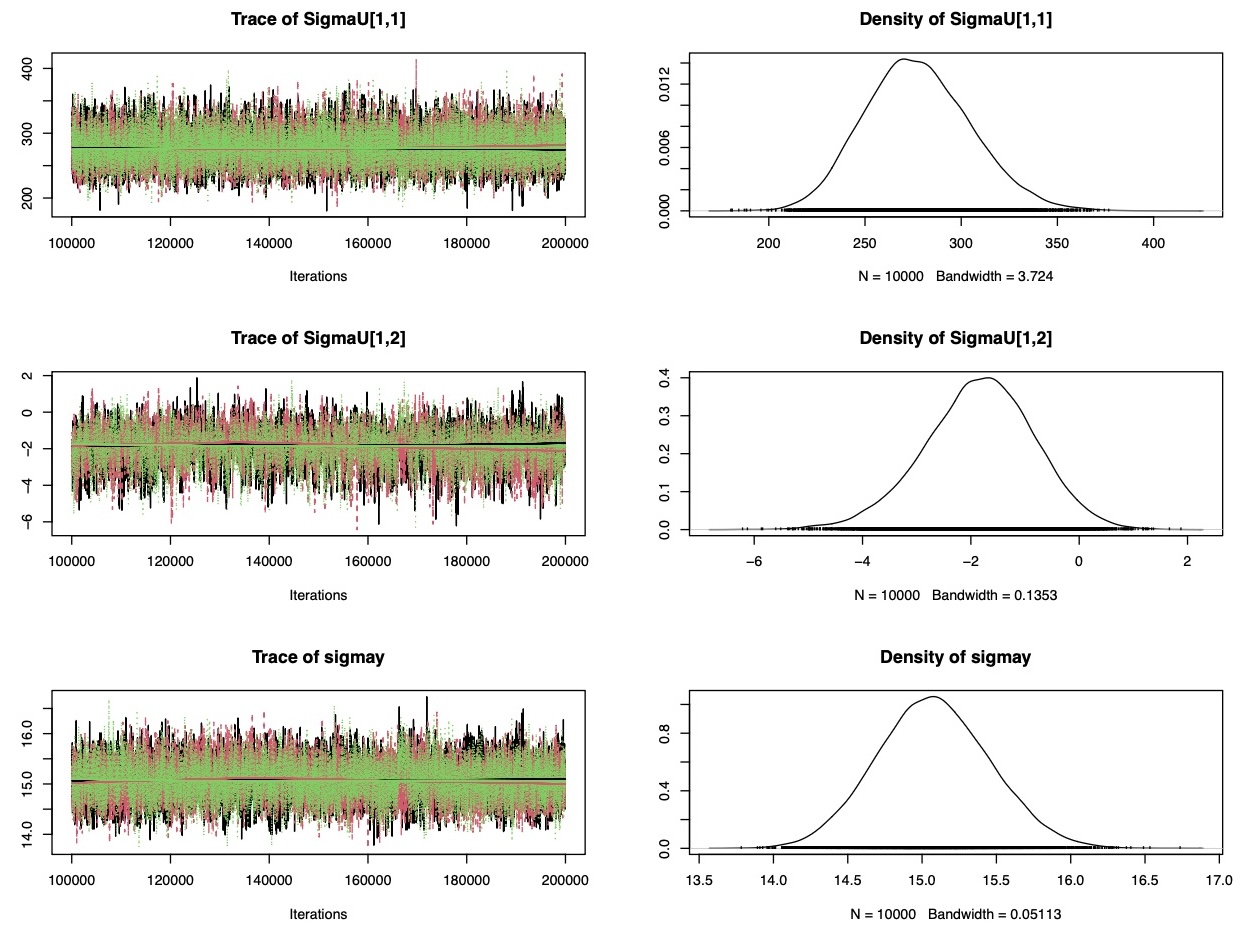}
    \caption{Traceplot and posterior density for parameters of joint model in the case study using data from one CV-fold with adequate convergence.}
    \label{trace_case1}
\end{figure}

\begin{figure}
    \centering
    \includegraphics[width=0.45\linewidth]{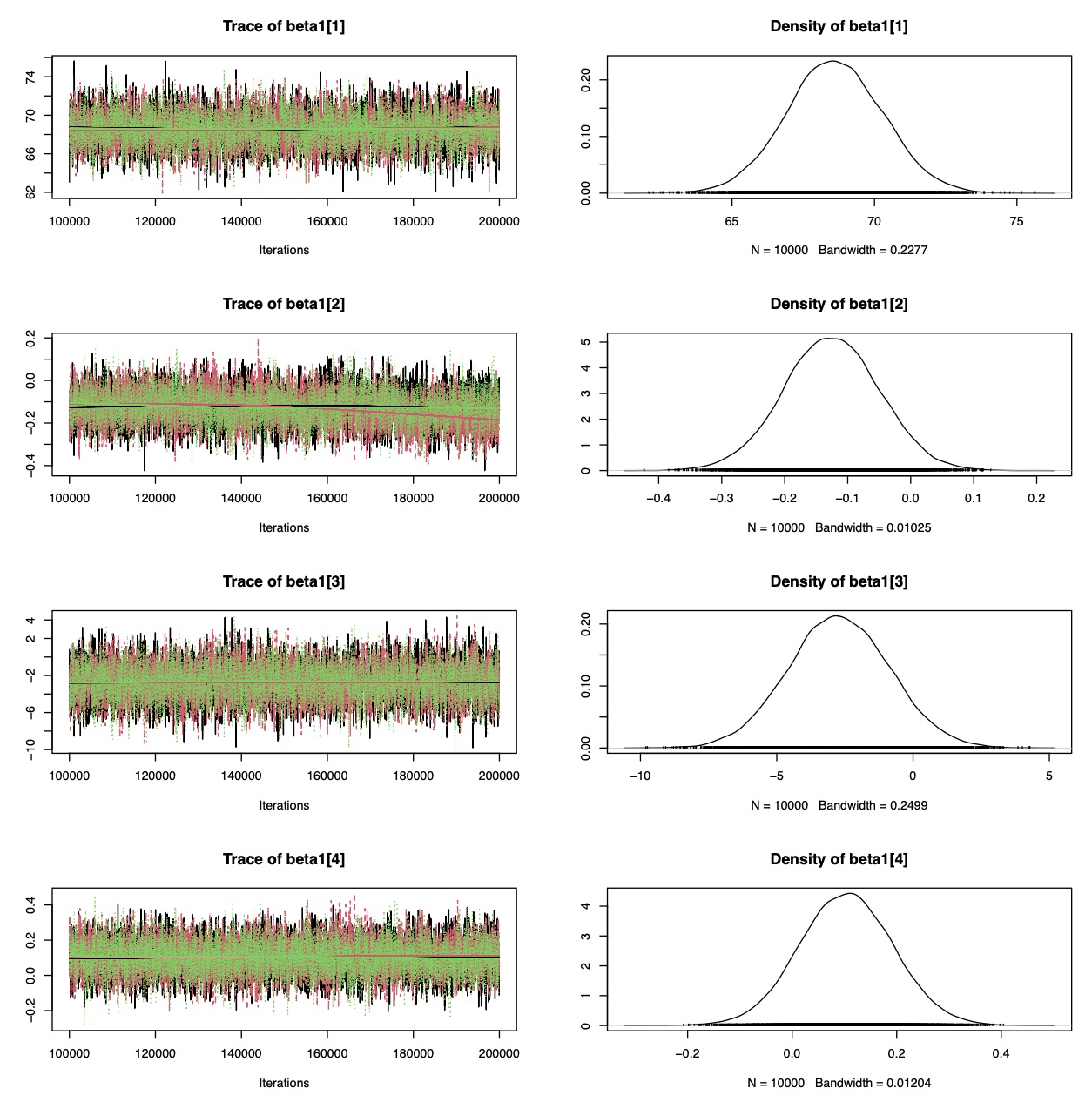}
     \includegraphics[width=0.45\linewidth]{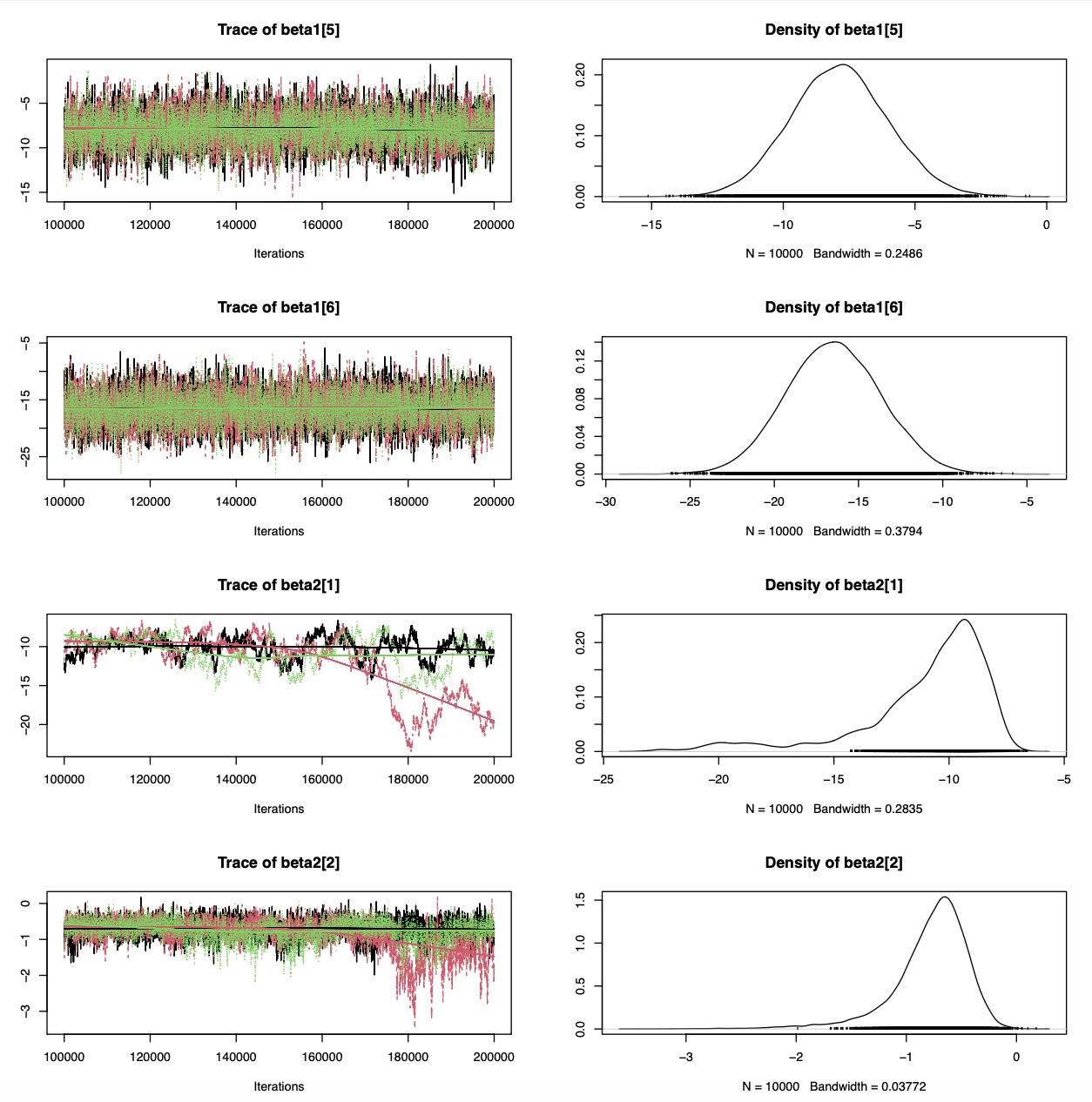}
      \includegraphics[width=0.45\linewidth]{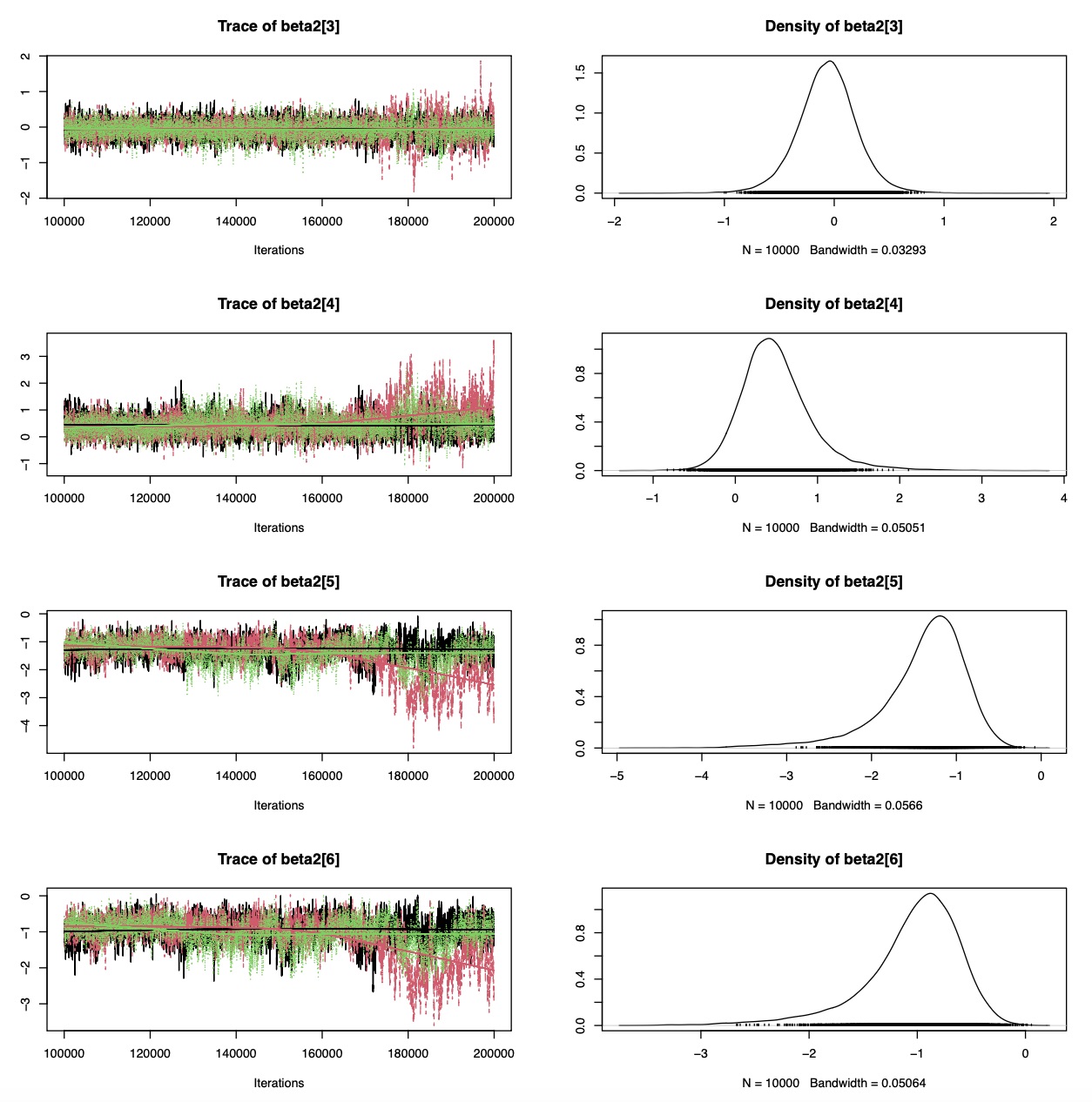}
       \includegraphics[width=0.45\linewidth]{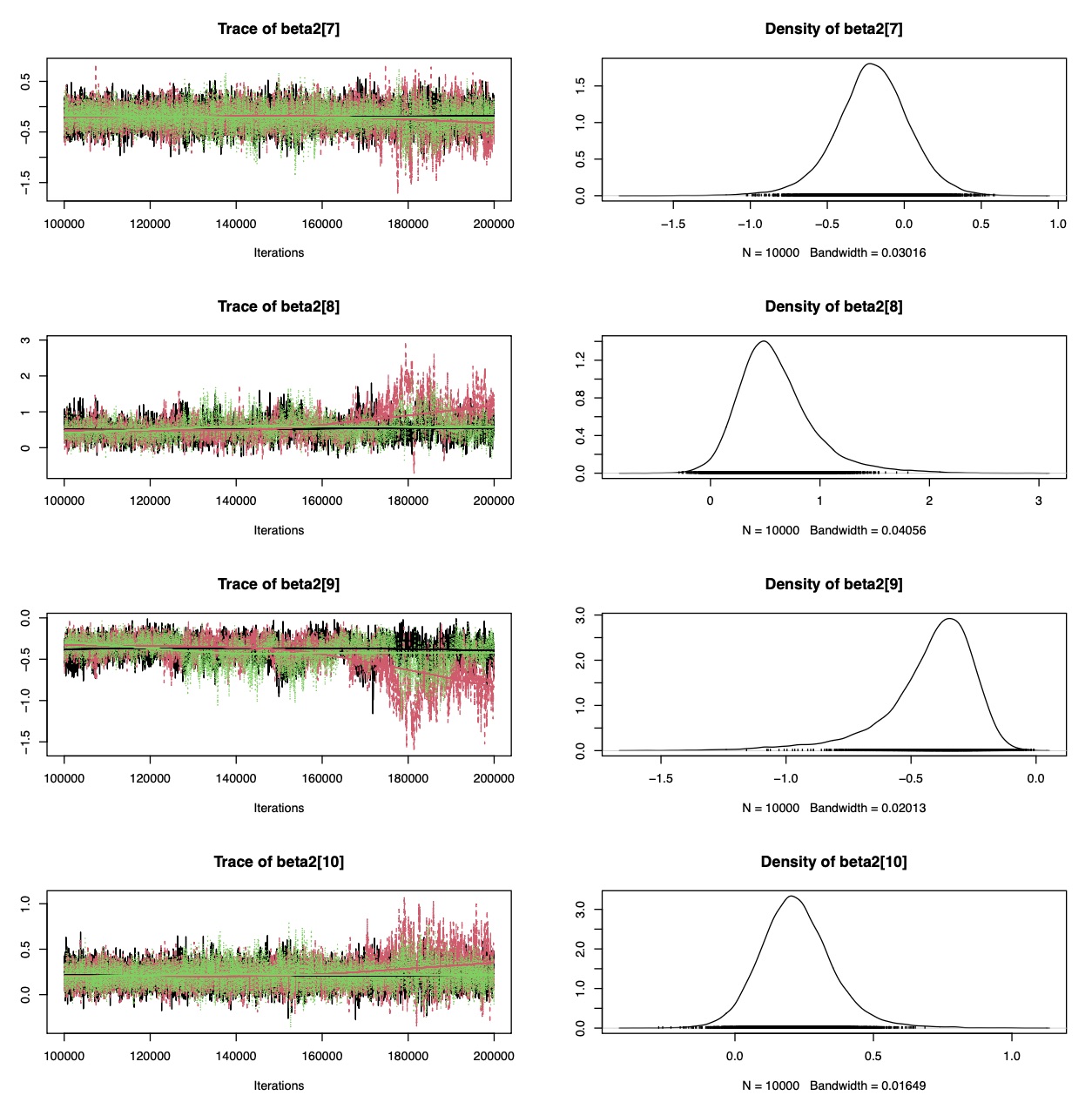}
        \includegraphics[width=0.45\linewidth]{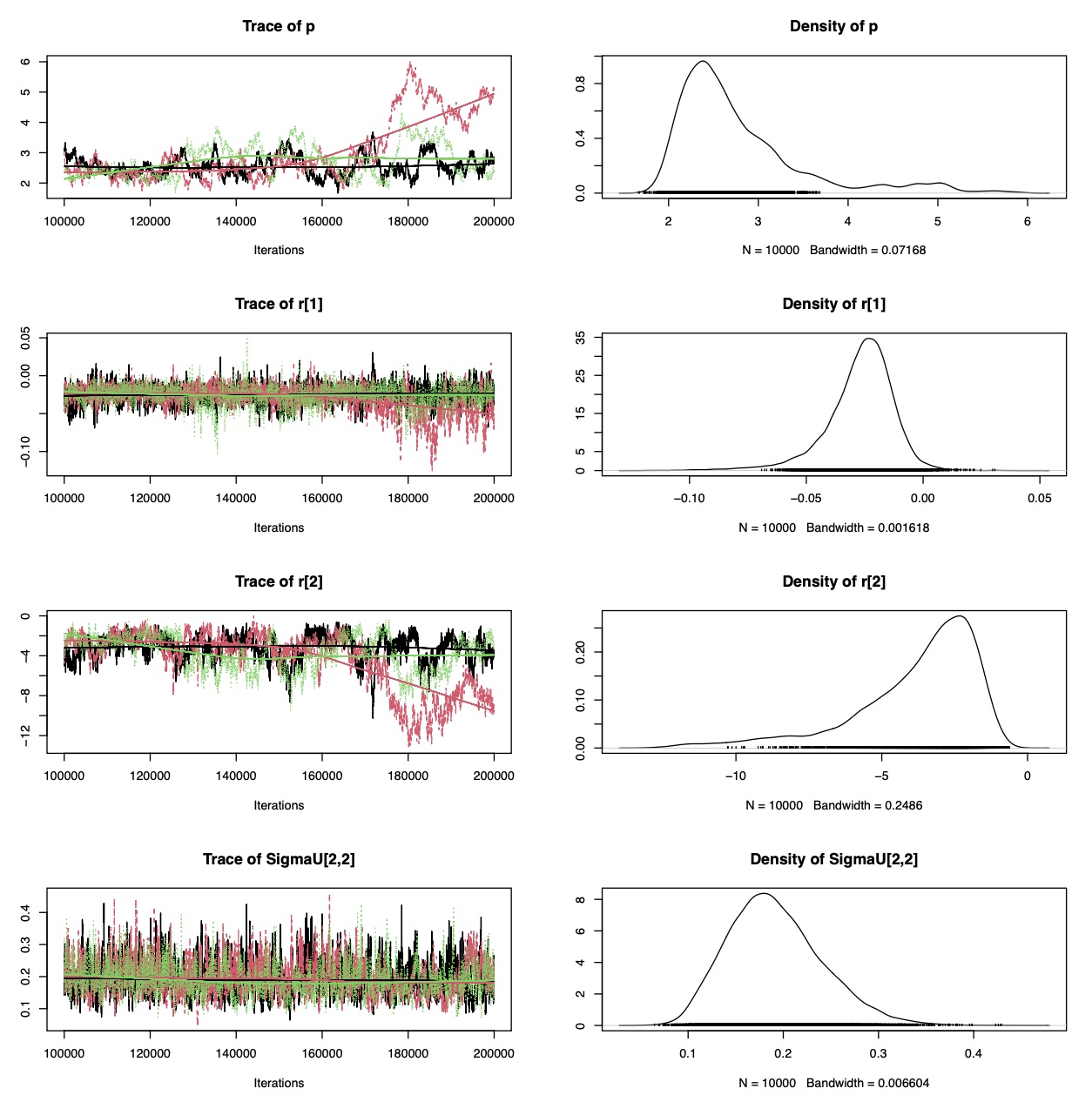}
         \includegraphics[width=0.45\linewidth]{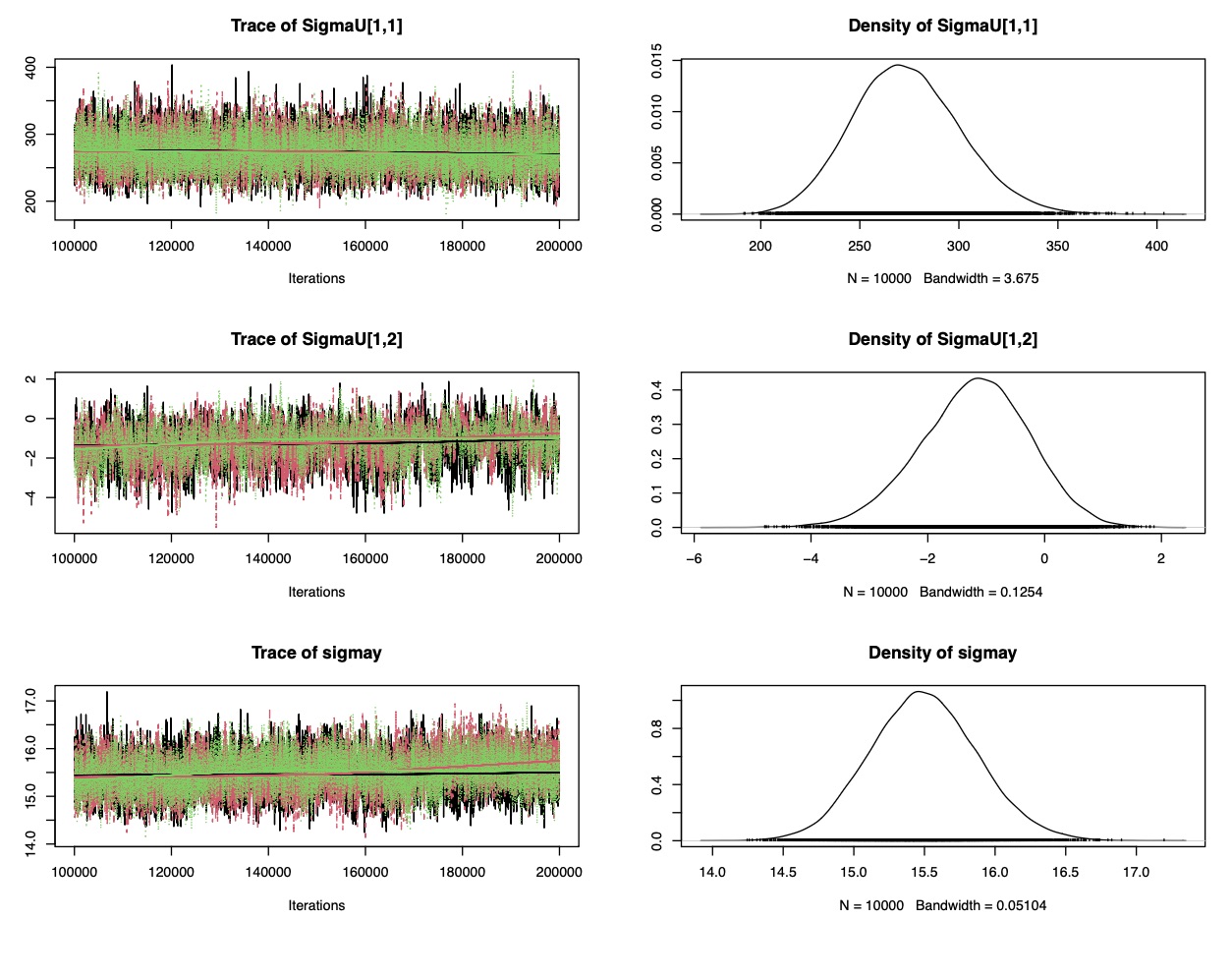}
   \caption{Traceplot and posterior density for parameters of joint model in the case study using data from the CV-fold with the poorest convergence.}
    \label{trace_case2}
\end{figure}

For the case study, the ESS for the longitudinal model parameters across the cross-validation folds ranged from \(2046\) to \(8573\), and from \(905\) to \(8934\) for the variance parameters. All Gelman–Rubin statistics for the longitudinal model were exactly \(1.00\), while the variance parameters reached a maximum of \(1.02\), indicating adequate convergence. For the survival model parameters, the ESS across folds ranged from \(33\) to \(16626\), with a maximum Gelman-Rubin statistic of \(1.18\). These results suggest that convergence was less satisfactory for some survival parameters. Inspection of the trace plots revealed that lack of global convergence was primarily observed for the Weibull shape parameter (\texttt{p}), the coefficient of the random slope (\texttt{r[2]}), and the intercept parameter (\texttt{beta2[1]}). In some instances, one Markov chain moved toward an alternative local optimum before occasionally returning. To mitigate the potential impact of this behavior on inference, we used the posterior median rather than the posterior mean as the point estimate for the survival parameters, as the median is less sensitive to multimodality and chain instability. Figures \ref{trace_case1} and \ref{trace_case2} display the trace plots and posterior densities for all parameters for two representative folds: one fold demonstrating adequate convergence (all Gelman-Rubin statistics $\leq 1.02$) and one fold exhibiting the poorest convergence for the survival parameters.

\begin{figure*}
\centering
\begin{verbatim}
model {
  for (i in 1:N) {
    # Longitudinal model
    for (j in 1:M) {
      Y[i,j] ~ dnorm(muy[i,j], tauy)
      muy[i,j] <- beta[i,1] + beta[i,2]*t[j] + beta1[3]*x[i] + beta1[4]*t[j]*x[i]
    }
    beta[i,1] <- beta1[1] + U[i,1]
    beta[i,2] <- beta1[2] + U[i,2]
     
    # Survival model
    is.censored[i] ~ dinterval(time[i],cen[i])
    time[i] ~ dweib(p, mut[i])
    log(mut[i]) <- beta2[1] + beta2[2]*x[i] +r[1]*U[i,1] + r[2]*U[i,2] 
    
    # Random effects
   U[i,1:2] ~ dmnorm(muU[], TauU[,])
  }

  #Priors
  ## Shape of Weibull
  p ~ dgamma(1,1)
  
  ## Variance y and random effects
  tauy ~ dgamma(0.01, 0.01)
  muU[1] <- 0
  muU[2] <- 0
  TauU[1:2,1:2] ~ dwish(R[,], 3)  
  R[1,1] <- 1
  R[1,2] <- 0
  R[2,1] <- 0
  R[2,2] <- 1
  SigmaU[1:2,1:2] <- inverse(TauU[,])
  sigmay <- tauy^(-1/2)
  
  ## Fixed effects
  beta1[1] ~ dnorm(0, 0.001)
  beta1[2] ~ dnorm(0, 0.001)
  beta1[3] ~ dnorm(0, 0.001)
  beta1[4] ~ dnorm(0, 0.001)
  beta2[1] ~ dnorm(0, 0.001)
  beta2[2] ~ dnorm(0, 0.001)
  r[1] ~ dnorm(0, 0.001)
  r[2]~ dnorm(0, 0.001)
}
\end{verbatim}
\caption{\texttt{BUGS} code for joint model in 2D simulation study}
\label{code2D}
\end{figure*}

\clearpage
\section*{Web Appendix C: Case study}
\subsection*{C1: Data descriptions}
\begin{figure}[h]
    \centering
    \includegraphics[width=\linewidth]{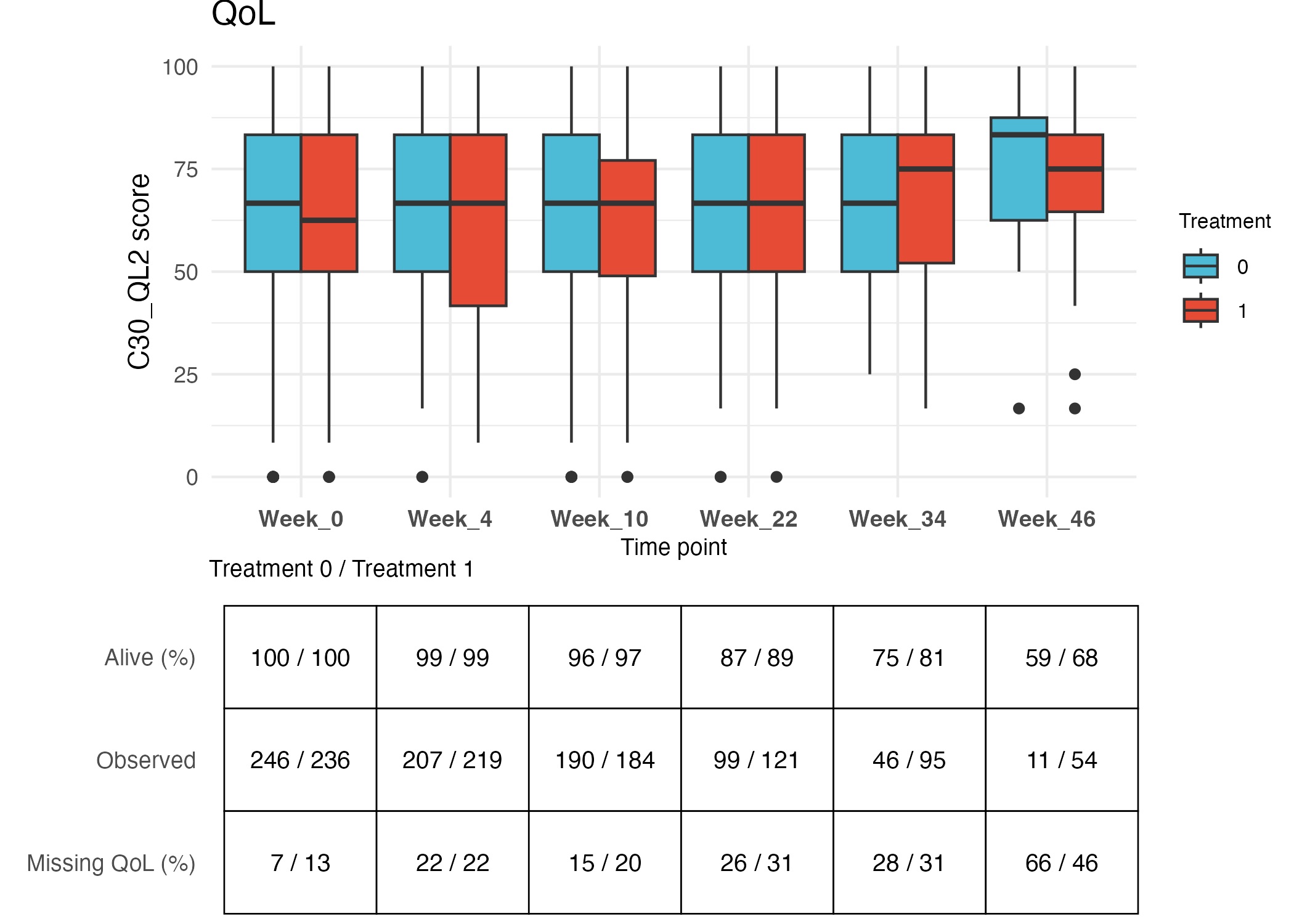}
    \caption{Overview of the data used in the case study. The upper panel displays boxplots of QoL measurements across time points, stratified by treatment group: radiotherapy alone (\(X = 0\) and radiotherapy plus temozolomide (\(X=1)\). The lower panel reports, for patients who initiated treatment, in row 1 the percentage of the patients observed to be alive, in row 2 the number of observed QoL measurements, and in row 3 the percentage of patients observed to have not progressed but did not report QoL at the considered time point.}
    \label{boxplot_data}
\end{figure}
\subsection*{C2: Positive and negative predictive value}
\begin{figure}
    \centering
    \includegraphics[width=\linewidth]{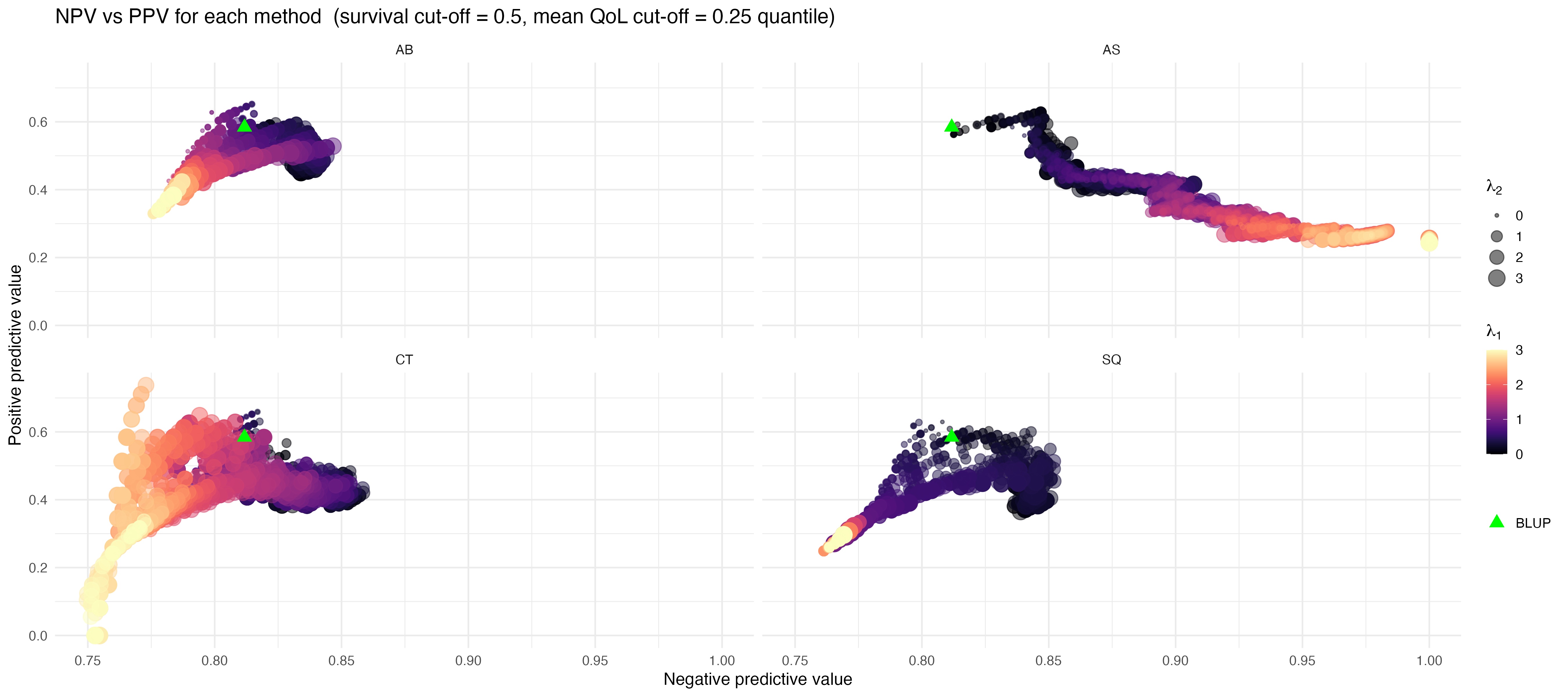}
    \includegraphics[width=\linewidth]{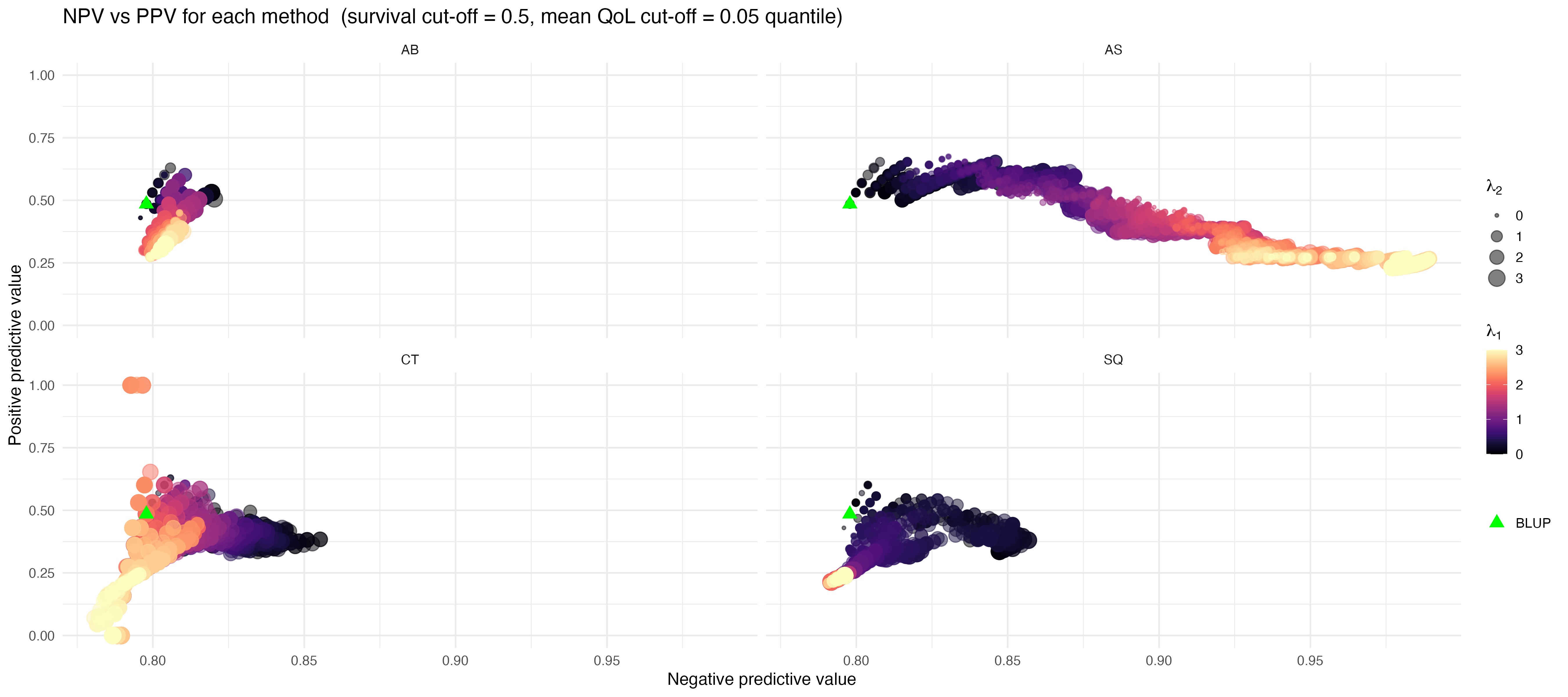}
    \caption{Positive Predictive Value (PPV) and Negative Predictive Value (NPV) for the case study, with outcomes evaluated over $t=(33,46)$. Upper panel: $c=0.5$ with the 25th percentile QoL threshold. Lower panel: $c=0.5$ with the 5th percentile threshold.}
    \label{flagging_data2}
\end{figure}

As an alternative to the true and false positive rates presented in the main manuscript, Figure \ref{flagging_data2} reports the positive and negative predictive values (PPV and NPV) for the WRaPs, based on a flagging rule defined by a predicted survival probability below 50\% and a predicted mean QoL below either the 25th or 5th percentile of the observed mean QoL in the dataset. The panels in these plots correspond to the different weight functions used to compute the WRaPs:
\begin{eqnarray*}
        W(u;\lambda)_{SQ} = \exp(\lambda u^2), 
        W(u;\lambda)_{AB} = \exp(\lambda |u|), \nonumber\\
        W(u;\lambda)_{CT} = I(|u| > \lambda), 
        W(u;\lambda)_{AS} = \exp(-\lambda u)
    \end{eqnarray*}
For the 25th percentile threshold, BLUPs yield a PPV of 58\% and a NPV of 81\%. Using WRaPs, with the AB weighting function, the PPV increases to 65\% while maintaining the same NPV. If a higher PPV is preferred over NPV, the CT weighting function achieves a PPV of 74\% with a slightly lower NPV of 77\%.
Under the 5th percentile threshold, the differences between BLUPs and WRaPs become more pronounced. BLUPs attain a PPV of 49\% and an NPV of 80\%. Both measures can be improved simultaneously using WRaPs: with the AS weighting function, a PPV of 67\% and an NPV of 83\% are obtained. Notably, the CT weighting function attains a PPV of 100\%, with only a slight reduction in NPV to 79\%.

\label{lastpage}

\end{document}